\documentclass[usegraphicx]{mn2e}
\usepackage{times}
\usepackage{graphicx,subfigure}

\begin{document}
\title[Dust in the Kepler and Tycho Remnants]{Dust in Historical
  Galactic Type Ia Supernova Remnants with {\it Herschel}\thanks{{\it
      Herschel} is an ESA space observatory with science instruments
    provided by European-led Principal Investigator consortia and with
    important participation from NASA.}}

\author[H.\ Gomez et al.] {H.\,L.\ Gomez $^{\!  1}$, C.\,J.\,R.\,Clark $^{\! 1}$,  T. Nozawa $^{\! 2}$, O. Krause $^{\!
    3}$, E.\,L.\,Gomez $^{\! 1,4}$, M. Matsuura $^{\!  5}$, \and M.\,J.\,Barlow
  $^{\!  5}$, M.-A. Besel $^{\!  3}$, L. Dunne $^{\!  6}$, W.\,K.\,Gear $^{\!  1}$, P. Hargrave $^{\!
    1}$, Th. Henning $^{\!  3}$, \and R.\,J.\ Ivison $^{7,8}$,  B.\
  Sibthorpe $^{\!  7}$, B.\,M.\,Swinyard $^{\! 5, 9}$, R. Wesson $^{\!
    10}$ \\
\\
  $^1$ School of Physics \& Astronomy, Cardiff University, The Parade,
  Cardiff, CF24 3AA, UK\\
  $^2$ Institute for the Physics and Mathematics of the Universe, University of Tokyo, Kashiwa, Chiba 277-8583, Japan\\
  $^3$ Max-Planck-Institut f\"{u}r Astronomie, K\"{o}nigstuhl 17,
  69117  Heidelberg, Germany \\
  $^4$ Las cumbres Obsveratory Global Telescope Network, 6740 Cortona
  Drive Suite 102, Goleta, CA 93117\\
  $^5$ Dept. of Physics and Astronomy, University College London,
  Gower Street, London WC1E 6BT, UK\\
  $^6$ School of Physics \& Astronomy, University of Nottingham,
  University Park, Nottingham NG7 2RD, UK\\
  $^7$ UK Astronomy Technology Centre, Royal Observatory
  Edinburgh, Blackford Hill, Edinburgh EH9 3HJ, UK\\
  $^8$ Institute for Astronomy, University of Edinburgh,
  Blackford Hill, Edinburgh EH9 3HJ, UK\\
  $^9$ Space Science and Technology Department, Rutherford Appleton Laboratory, Oxfordshire, OX11 0QX, UK \\
  $^{10}$ European Southern Observatory, Alonso de Cordova 3107, Casilla 19001, Vitacura, Santiago 19, Chile
}
\date{}

\pagerange{\pageref{firstpage}--\pageref{lastpage}} \pubyear{2011}

\maketitle
 
\begin{abstract}
  The origin of interstellar dust in galaxies is poorly understood,
  particularly the relative contributions from supernovae and the cool
  stellar winds of low-intermediate mass stars. Recently, large masses
  of newly-formed dust have been discovered in the ejecta of
  core-collapse supernovae.  Here, we present {\it Herschel} PACS and
  SPIRE photometry at 70-500\,$\mu$m of the historical, young
  supernova remnants: Kepler and Tycho; both thought to be the
  remnants of Type Ia explosion events. We detect a warm dust
  component in Kepler's remnant with $T_d =82^{+4}_{-6} \, \rm K$ and
  mass $\rm \sim (3.1^{+0.8}_{-0.6})\times 10^{-3} \,M_\odot$; this is
  spatially coincident with thermal X-ray emission and optical knots
  and filaments, consistent with the warm dust originating in the {\it
    circumstellar} material swept up by the primary blast wave of the
  remnant.  Similarly for Tycho's remnant, we detect warm dust at
  $90^{+5}_{-7}\,\rm K$ with mass $(8.6^{+2.3}_{-1.8})\times
  10^{-3}\,\rm M_{\odot}$.  Comparing the spatial distribution of the
  warm dust with X-rays from the ejecta and swept-up medium, and
  H$\alpha$ emission arising from the post-shock edge, we show that
  the warm dust is swept up {\it interstellar} material.  We find no
  evidence of a cool (25-50\,K) component of dust with mass $\ge
  0.07\,\rm M_{\odot}$ as observed in core-collapse remnants of
  massive stars.  Neither the warm or cold dust components detected
  here are spatially coincident with supernova ejecta material.  We
  compare the lack of observed supernova dust with a theoretical model
  of dust formation in Type Ia remnants which
  predicts dust masses of $88(17) \times 10^{-3} \,\rm M_{\odot}$ for
  ejecta expanding into ambient surrounding densities of $1(5)\,\rm
  cm^{-3}$. The model predicts that silicon- and carbon-rich dust
  grains will encounter, at most, the interior edge of the observed
  dust emission at $\sim$400\,years confirming that the majority of the warm dust originates from swept up circumstellar or interstellar
  grains (for Kepler and Tycho respectively).   The lack of cold dust grains in the ejecta suggests that Type Ia remnants do not produce substantial quantities of iron-rich dust grains and has important consequences for the `missing' iron mass observed in ejecta.  Finally, although, we cannot
  completely rule out a small mass of freshly-formed supernova dust,
  the {\it Herschel} observations confirm that significantly less dust forms in the
  ejecta of Type Ia supernovae than in the remnants of core-collapse
  explosions.
\end{abstract}


\section{Introduction}
Dust in galaxies is thought to be produced by in the stellar winds of
both low-intermediate mass (LIM) Asymptotic Giant Branch (AGB) stars
(e.g. Gehrz 1989; Whittet 2003; Sargent et al. 2010) and, to an
unknown extent, by massive stars (H\"{o}fner 2009; Gomez et al.\ 2010;
Andrews et al. 2011; Gall, Hjorth \& Anderson 2011a). Massive stars may
also produce dust in the expanding gas ejecta when they explode as
supernovae (e.g.  Clayton et al.\ 2001; Todini \& Ferrara 2001). In
the Milky Way (see Zhukovska \& Gail 2008), the major dust source is
presumed to be LIM stars, but when accounting for dust destruction
timescales and the observed dust mass in the Galaxy, the dust
injection rate from stars required is an order of magnitude higher
than observed. An alternative source of dust is required to make up
the dust budget (e.g. Jones 2001). This shortfall in the dust mass
estimated from the dust injection rates from LIM stars is also
observed in the interstellar medium (ISM) of the Large Magellanic
Cloud (Matsuura et al.\ 2009), in galaxies out to $z=0.4$ (Dunne et
al.\ 2011) and in dusty high-redshift galaxies (see Morgan \& Edmunds
2003; Dwek, Galliano \& Jones \ 2007).  Although recent work by Jones
\& Nuth (2011) suggest that the dust-budget problem is not as
problematic as previous works suggest given the uncertainties in mass
loss rates and the efficiencies of dust formation in stellar winds, it
is clear that significant dust production in SN ejecta would alleviate
this dust budgetary problem.  The average dust yield per SN required
to explain dusty galaxies at both low and high redshifts is of order
$0.5-1\,\rm M_{\odot}$ (Micha{\l}owski, Watson \& Hjorth 2010; Gall,
Anderson \& Hjorth 2011b).

  Although SNe have long been proposed as a source of dust in the ISM
  (Dwek \& Scalo 1980; Clayton et al.\ 2001), observational evidence
  has been scarce until recently. Wooden et al.\ (1993) first detected
  emission from $10^{-4}\,\rm M_{\odot}$ of dust in SN1987A using the
  Kuiper Airborne Observatory.  Subsequent Mid-far-infrared (FIR) observations of
  Galactic and extra-galactic core-collapse remnants with the {\it
    Spitzer} Space Telescope detect small quantities of warm dust
  ($\sim (1-50) \times 10^{-3}\,\rm M_{\odot}$) in young
  ($t<$1000\,days - Sugerman et al.\ 2006; Meikle et al.\ 2007, 2011;
  Kotak et al.\ 2009; Andrews et al.\ 2010, 2011; Fabbri et al 2011;
  Szalai et al.\ 2011) and old remnants (Williams, Chu \& Gruendl
  2006; Rho et al.\ 2008).  The evidence for dust production in Type Ia
  remnants is scarce (Borkowski et al.\ 2006; Seok et al.\ 2008):
  mid-IR observations of the Kepler SNR with the {\it Infrared Space
    Observatory} ({\it ISO}; Douvion, Lagage \& Pantin 2001) found
  thermal emission from warm dust (100\,K) with mass $10^{-4}\,\rm
  M_{\odot}$. More recently, Ishihara et al.\ (2010) detected
  $10^{-3}\,\rm M_{\odot}$ of swept-up (warm) dust in the Tycho SNR
  with {\it AKARI}, with a tentative suggestion that $10^{-4}\,\rm
  M_{\odot}$ of dust seen in the north-east of the remnant may have
  been formed in the ejecta. Outside of our Galaxy, Borkowski et al.\
  (2006) obtained mid-IR observations of four Ia remnants in the Large
  Magellanic Cloud, finding $<10^{-2}\,\rm M_{\odot}$ of swept-up
  interstellar dust. The dust masses from the mid-far IR studies are
  orders of magnitude lower than needed to solve the dust budget
  issues.

  In order to investigate if supernova remnants (SNRs) contain cold
  dust missed by previous infrared (IR) observations, the
  Submillimetre Common User Bolometer Array (SCUBA) was used to
  observe Cassiopeia~A (Cas A), a remnant of a Galactic core-collapse SNe
  (Krause et al.\ 2008a). The cold dust seen at 450 and 850\,$\mu$m
  was interpreted as emission from dust associated with the remnant
  due to the spatial correlation with the X-ray, radio and
  submillimetre (submm) emission (Dunne et al.\ 2003). Subsequently, a
  significant fraction of the submm emission was shown to originate
  from molecular clouds along the line of sight (Krause et al.\ 2004;
  Wilson \& Batrla 2005). Recent observations with the Balloon-borne
  Large Aperture Submillimeter Telescope, BLAST and the {\it Herschel}
  Space Observatory showed it was difficult to distinguish between
  cold dust in the remnant and cold dust from intervening interstellar
  clouds using photometry alone (Sibthorpe et al.\ 2010; Barlow et
  al.\ 2010). These studies revealed a new cool component of dust in
  Cas A with mass $0.08\,\rm M_{\odot}$ at $\sim 35\,\rm K$, yet even
  if all of the dust survived the passage through the shock, the dust
  mass is about one order of magnitude lower than necessary to solve
  the dust budget problem.

A possible method to distinguish between supernova dust and
contamination from interstellar material along the line of sight was
proposed by Dunne et al.\ (2009) where they traced the alignment of
the dust towards Cas A with the SNR magnetic field using polarimetry
at 850\,$\mu$m. They find that a significant fraction of the submm
emission towards Cas A is strongly aligned with the magnetic field of
the remnant. The dust mass inferred from the polarimetry study of Cas
A agrees with the dust masses estimated from {\it Herschel} photometry
observations of the core-collapse SNe SN1987A (Matsuura et al.\
2011); in the
former $0.2-1.0\,\rm M_{\odot}$ of dust was detected depending on the
composition. The limit on the ejected heavy element mass in these
sources places a further constraint on the maximum physical dust mass,
finding $0.2<M_d <0.5\,\rm M_{\odot}$.  

The evidence for dust formation in
core-collapse SNRs is accumulating, though it is not yet
understood whether the amount of SN dust created is dependent on the
progenitor mass, the metallicity or the type of explosion. Neither is it clear
how much dust will survive, though theoretical studies suggest a
significant fraction of dust grains formed in ejecta will be destroyed
before injection into the ISM (Jones 2001; Bianchi \& Schneider 2007)
particularly in the ejecta of envelope-poor core-collapse SNe or Type
Ia remnants (Nozawa et al.\ 2010; Nozawa et al.\ 2011). As yet, there
is no clear evidence for freshly formed dust in the remnants of Type
Ia explosions, though theoretical models of dust formation in the
remnants of Type Ia events suggests that up to $0.2\,\rm M_{\odot}$
dust could form in the ejecta (Nozawa et al.\ 2011).

Historical remnants are unique probes for understanding dust
production and indeed, destruction following core collapse Type II
supernovae as well as low mass binary Type I events. In the {\it
  Herschel} era, we are finally beginning to probe the contribution of
SNe to the dust budget, yet given the somewhat large uncertainties in
our understanding of the dust yield from stellar sources in other
galaxies and in interpreting statistical descriptions of the dust content
in galaxies from wide-area surveys (e.g. Dunne et al. 2011), it is
important to investigate dust production in those nearest, youngest
and resolved galactic remnants. One can also argue that understanding
dust formation in supernovae is important regardless of the explosion
mechanism: dust formation following Type-II SNe might suggest that
Type-Ia SNe would also be likely dust producers (Clayton et al.\ 1997;
Travaglio et al.\ 1999). Alternatively, detecting dust from
circumstellar material swept up by the blast wave of a Type Ia can be
a useful probe of the progenitor and explosion mechanism.

The resolution, wavelength coverage and sensitivity of {\it Herschel}
allows us to investigate the origin and
quantity of dust in historical supernova remnants.  Here we present IR
and submm observations (\S\ref{sec:obs}) of the Tycho and Kepler
remnants.  We determine the dust mass in and towards the remnants in
\S\ref{sec:kepler} and \S\ref{sec:tycho} and compare the IR emission
with multiwavelength tracers in \S\ref{sec:resultskepler} and
\S\ref{sec:resultstycho}. The origin of the dust is discussed in
\S\ref{sec:disc} and we use the theoretical model of Nozawa et al.\
(2011) to compare the expected dust masses with the observations.  The
results are summarised in \S\ref{sec:conc}.

\section{Observations and data reduction}
\label{sec:obs}

\subsection{Herschel Photometry}
The Kepler and Tycho supernova remnants were observed with the {\it
  Herschel} (Pilbratt et al.\ 2010) Photodetector Array Camera and Spectrometer (PACS;
Poglitsch et al.\ 2010) and Spectral and Photometric Imaging Receiver
(SPIRE; Griffin et al. 2010) at 70, 100, 160, 250, 350 and 500\,$\mu$m
as part of the Mass Loss from Evolved StarS survey (MESS: Groenewegen
et al.\ 2011). Kepler was observed on 10th Sept and 11th March 2010;
Tycho on 20th March 2010 and 18th Jan 2010 for PACS and SPIRE
respectively. The PACS data were obtained in `scan-map' mode with
speed $20^{\prime \prime}\rm /s$ including a pair of orthogonal
cross-scans over an area $\sim 22^{\prime} \times 22^{\prime}$. The
SPIRE maps were observed in `Large Map' mode with scan length of
$30^{\prime}$ over $\sim 32^{\prime} \times 32^{\prime}$; a cross-scan
length is also taken, with a repetition factor of three. The data were
processed following the detailed description given in Groenewegen et
al.\ (2011).

The PACS maps were reduced with the {\it Herschel}
Interactive Processing Environment ({\small HIPE}; Ott 2010) applying all low-level
reduction steps (including deglitching) to Level 1. The {\small SCANAMORPHOS}
software (Roussel 2011) was then used remove effects due to the thermal
drifts and the uncorrelated $1/f$ noise of the individual
bolometers. Final Level 2 map creation through data projection onto a
changeable spatial grid was performed with {\small SCANAMORPHOS}. Since the
{\sc HIPE} reduction was based on the calibration file version v5,
interim calibration factors of 1.12, 1.15 and 1.17 were applied to the
PACS data at the respective wavelengths to be consistent with the
current calibration file v6.  The Full Width Half Maximum ({\sc fwhm}) at 70, 100 and
160\,$\mu\rm m$ is 6, 8 and 12$\arcsec$ respectively.  The flux
calibration uncertainty for PACS is currently estimated as
10\,per\,cent for the 70 and
100\,$\mu$m bands and 20\,per\,cent at 160\,$\mu$m (Poglitsch et al.\
2010). Colour corrections for PACS are expected to be small compared
to the calibration errors.  To correct for colour with dust spectrum
$\nu^{1.0}$ and temperatures of $100\,\rm K$, we divided by 1.00 and
1.03\footnote{\em
  http://herschel.esac.esa.int/twiki/pub/Public/PacsCalibrationWeb/}
at 70 and 100\,$\mu$m. The 20\,K interstellar dust component dominates
the 160\,$\mu$m image, and as such the correction factor is 0.957.  

For SPIRE, the standard photometer pipeline ({\small HIPE} v.5.0) was
used (Griffin et al.\ 2010) with an additional iterative baseline
removal step (e.g. Bendo et al.\ 2010). The SPIRE maps were created
with the standard pipeline {\small NA\"IVE} mapper.  We multiply the
350\,$\mu$m data product by 1.0067 to be in line with most recent
calibration pipeline (equivalent to {\small HIPE} v7 reduction). Since
the beam areas increase as a function of pixel scale\footnote{\em
  http://herschel.esac.esa.int/Docs/SPIRE}, the {\sc fwhm} is 18.1,
24.9 and 36.4$^{\prime \prime}$ for pixel sizes of 6, 10 and
14$^{\prime
  \prime}$, at 250, 350 and 500\,$\mu\rm m$ respectively. The SPIRE
calibration methods and accuracies are outlined by Swinyard et al.\
(2010) and are estimated to be approx 10\,per\,cent. The pipeline
produces monochromatic flux densities for point sources with $\nu
S_{\nu} \propto \nu^{1.0}$ but at the longer wavelengths, colour
corrections become significant. We multiplied by the colour corrections $
0.999, 1.0045, 1.0302$ as defined in the {\it
  Herschel} SPIRE manual$^{2}$, appropriate for $\beta=1.5$ and
extended sources.

\subsection{Multiwavelength Datasets}

Optical data for the Tycho and Kepler remnants were obtained from the {\it Hubble Space
  Telescope} ({\it HST}) archive via {\sc
  ALADIN} (Bonnarel et al.\ 2000). For Kepler's remnant we used
observations in the filters 658, 660 and 502$\rm \,nm$ where the
emission originates from $\rm H\alpha$ + [N{\sc ii}], [N{\sc ii}] and
[O{\sc iii}] respectively (e.g. Sankrit et al.\ 2008). Bad pixels were
masked by hand. For Tycho, we use data observed at 656\,nm (Lee et
al.\ 2010), tracing $\rm H\alpha$ in the eastern limb of the remnant.

{\it Spitzer} Multiband Imaging Photometer (MIPS; Rieke et al. 2004)
Level 2 calibrated data were obtained via the archive\footnote{\em
  http://irsa.ipac.caltech.edu/data/SPITZER/docs/spitzerdataarchives/}
(Kepler: PI W.Blair; Tycho: PI J.Rho) at 24 and 70\,$\mu$m. For
100\,$\mu$m data, we use the Improved Reprocessing of the {\it IRAS}
Survey (IRIS; Milville-Deschenes \& Lagache 2005) which have been
calibrated with respect to data taken with the Diffuse Infrared
Background Experiment (DIRBE).  Wide field Infrared Survey Explorer
(WISE; Wright et al.\ 2010) data were also obtained at 22\,$\mu$m.

The radio data at 18\,cm for Tycho was obtained through the Very Large
Array (VLA) archive. The Kepler data at 6\,cm was kindly provided by
T.\,DeLaney (DeLaney et al.\ 2002).  The {\it Chandra} X-ray data for
Kepler (PI S. Reynolds; Reynolds et al.\ 2008) and Tycho (PI J.Hughes;
see Badenes et al.\ 2006; Katsuda et al.\ 2010) were available online
as calibrated files extracted into soft and hard regions of the
X-ray spectrum\footnote{\em
  http://chandra.harvard.edu/pwarmo/openFITS/}. For Kepler's remnant,
the X-ray data used here range from $0.30-0.72$, $0.72-0.90$ and
$1.7-2.1\,\rm keV$, and for Tycho, we have $0.95-1.26$, $1.63-2.26$
and $4.1-6.1\,\rm keV$.  These ranges show line emission from silicon
and iron orginating from the ejecta as well as soft and hard
(with thermal and non-thermal contributions) X-ray continuum
orginating from the blast wave.

To compare the submm data with the molecular emission towards and
surrounding the remnants, we use high resolution $^{12}{\rm
  CO}(J=2-1)$ and $^{13}{\rm CO}(J=2-1)$ observations for Kepler (see
Gomez et al.\ 2009 for full details).  We also include our SCUBA image
of Kepler at 850\,$\mu$m (Morgan et al.\ 2003; Gomez et al.\ 2009).
Low resolution carbon monoxide $^{12}{\rm CO}(J=1-0)$maps for
Tycho's remnant were taken from the Canadian Galactic Plane Survey
(CGPS; Taylor et al.\ 2003). The data were integrated over $-68$ to
$-53\rm \,km\,s^{-1}$ to compare the remnant with molecular gas
thought to be interacting with the supernova remnant (see
\S\ref{sec:resultskepler}).

\subsection{Zero Levels}
\label{sec:zero}
\begin{figure}
   \centering 
\includegraphics[trim=5mm 14mm 12mm 0mm,clip=true,width=8cm]{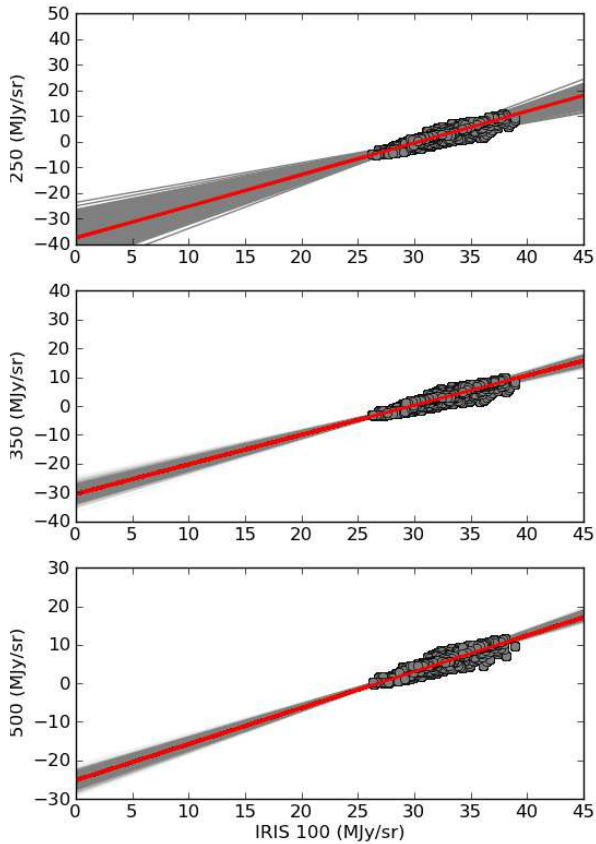}
\caption{\small{The {\it Herschel} SPIRE intensities for the Kepler
  observations versus {\it IRAS} IRIS 100-$\mu$m in MJy/sr.  The SPIRE maps
  are convolved to the same resolution as the IRIS maps and regridded
  onto the same pixel scale.  The red line is the least squares
  function, the grey circles are the {\it Herschel} data divided by
  the gain factor (due to the colour correction between IRIS at
  100\,$\mu$m and the SPIRE wavelengths).  The data are masked at the
  location of the remnant.  The grey shaded regions indicate the
  results from the 1000 bootstrap fits to the same dataset.  }}
         \label{fig:zero}
   \end{figure}

Due to the difficulty in finding a suitable baseline with which to
subtract the background in the {\it Herschel} SPIRE maps, we correlate
the pixel intensity from {\it Herschel} with the archival {\it IRAS}
IRIS maps (calibrated with respect to DIRBE) to estimate the zero
level (see Fig.~\ref{fig:zero}, Table~\ref{tab:zero}).  The {\it
  Herschel} data were convolved to the same angular resolution as the
IRIS 100-$\mu$m map ($258^{\prime \prime}$) and regridded to
$90^{\prime \prime}$ pixels. A least squares fit analysis was
performed on the {\it Herschel} vs. IRIS data (whilst masking out the
remnant) to determine the flux offsets; this sets the zero level of
the PACS and SPIRE data assuming that the IRIS data is a true
representation of the sky. A further scaling factor needs to be
accounted for due to the expected colour correction between the bands
when comparing to the 100\,$\mu$m band (the so-called `gain'
factor). These were estimated by interpolating a modified blackbody
curve $\nu^2B({\nu},T_d)$ (after a preliminary correction of the zero
levels). The zero level of the PACS and SPIRE maps are negative except
for the PACS 70\,$\mu$m map of Kepler. The least squares fit was then
repeated (see e.g. Bracco et al.\ 2011). The errors on the fitting
were estimated using the confidence intervals from the bootstrap
method, where 1000 fits were made to the bootstapped data,
Table~\ref{tab:zero}. This error is added in quadrature along with the
calibration errors listed above and introduces (at most) an error of
$\sim 17\%$ (well within the 20\% calibration error used for PACS at
160\,$\mu$m); note that this does not account for any systematic
errors.

\begin{table}
\begin{center}
  \begin{tabular}{ccc}\\ \hline \hline
    \multicolumn{1}{c}{Wavelength }&
    \multicolumn{2}{c}{Zero level ($\rm MJy/sr$)} \\ 

    \multicolumn{1}{c}{ ($\rm \mu m$)}&
    \multicolumn{1}{c}{Kepler} &\multicolumn{1}{c}{Tycho}\\ \hline
     70  & $4.7\pm 0.4$ & $-0.9\pm 0.1$\\
     100  & $-3.1\pm 0.2$ & $-2.5\pm 0.3$\\
     160   & $-4.8\pm 0.3$ &$-2.6\pm 0.2$ \\
    250   & $-37.4 \pm 3.8$ &$-30.8\pm 4.8$ \\
    350 & $-30.5 \pm 2.0$& $-20.4\pm 2.1$  \\
    500  & $-25.2 \pm 1.4$&$-7.7\pm 1.3$ \\ \hline
\end{tabular}
\end{center}
\caption{\small{Zero level on the PACS and SPIRE maps calculated using least squares fit to {\it
    Herschel} data and {\it IRAS} IRIS data.  Errors are the 1\,$\sigma$ estimated
  from bootstrapping the data 1000 times and refitting each set.}}
\label{tab:zero}
\end{table}

\section{Kepler's Supernova Remnant}
\label{sec:kepler} 

The Kepler supernova event was first seen by Johannes Kepler in 1604
and the remnant has a shell-like structure $\sim$3\,arcmin in
diameter. Estimates of its distance, using H\,I absorption features,
range from 3.9 to 6\,kpc (Reynoso \& Goss 1999; Sankrit et al.\ 2008).
Its classification has been controversial (see the excellent reviews
in Blair et al.\ 2007; Reynolds et al.\ 2008; Sankrit et al.\ 2008)
with evidence pointing towards both Type-Ia -- the thermonuclear
explosion of a low-mass accreting star in a binary system -- or 
Type-Ib -- the core collapse of a massive star. The deepest X-ray
analysis with {\it Chandra} (Reynolds et al.\ 2008), however, suggests that
a Type-Ia is the most likely classification using both iron abundance
constraints and the lack of a neutron star. The deep X-ray images also
provide evidence of the blast-wave interacting with dense
circumstellar material (CSM). This is unexpected for Type Ia events
although observations of CSM are becoming more widespread; the
presence of such a dense CSM surrounding Kepler led Reynolds to
suggest the thermonuclear explosion of a single massive star ($M_{*}
\sim 8\,\rm M_{\odot}$) as an alternative to the canonical White-Dwarf
(WD) binary scenario of Ia explosions. Whether Kepler was the result
of a single progenitor or WD-AGB-binary system explosion, the
mechanism was likely to be a deflagration explosion (Yang et al.\
2009; see Woosley, Taam \& Weaver 1986 for a review on explosion models). In
this scenario, the heating of the ejecta by radioactive elements is
higher than in core-collapse remnants, the ejecta densities will also
decrease rapidly, though the ejecta is well mixed since the reverse
shock progresses quickly through the layers in a deflagration
explosion (see \S\ref{sec:model}).

\begin{figure}
   \centering 
\includegraphics[trim=10mm 0mm 25mm
   0mm,clip=true,width=8.5cm]{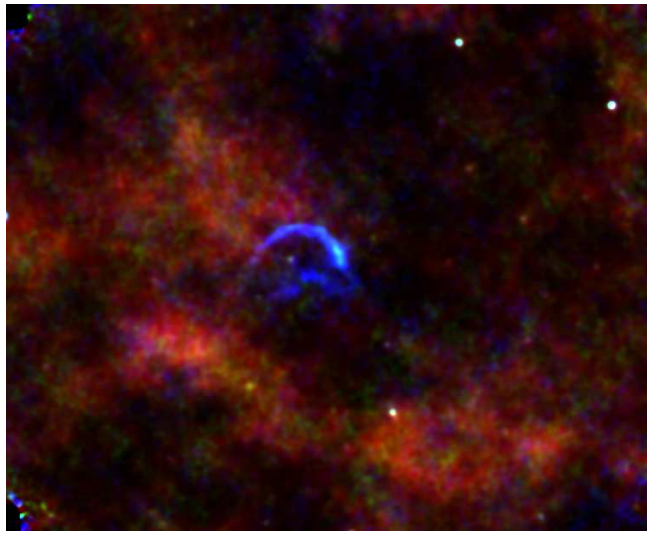}
   \caption{\small{Three-colour FIR-submm image of Kepler's SNR with
       PACS at 70 (blue), 100 (green) and 160\,$\mu$m (red).  The
       image has been smoothed.  The area shown is $16.5^{\prime}
       \times 16.6^{\prime}$. }}
         \label{fig:3colkep}
   \end{figure}

The warm dust emission detected in Kepler with {\it ISO} (Douvion et
al.\ 2001), follows the well-known north-south shell asymmetry seen
also at optical, X-ray and radio wavelengths. The correlation of the
dust with the optical $\rm H{\alpha}$ suggested the dust was
shock-heated circumstellar dust, swept up by the shockwave of the blast.
Blair et al.\ (2007) used {\it
  Spitzer} to revise the dust mass to $\rm 5.4\times
10^{-4}\,M_{\odot}$. They also found that emission at $70\,
\mu\rm m$ arises only in the brightest knots of the shell seen at
$24\,\rm \mu m$.  The $160\,\rm \mu m$ {\it Spitzer} data was
poor-quality due to its low resolution and strong background gradient.
Morgan et al.\ (2003) originally interpreted their longer wavelength 450
and 850-$\rm \mu m$ SCUBA data as evidence for
0.3--1\,$\rm{M_{\odot}}$ of cold dust associated with the remnant.
They compared the emission from interstellar material towards Kepler
using submm continuum imaging and spectroscopic observations of atomic
and molecular gas, via H\,I, $^{12}$CO and $^{13}$CO and
detected weak CO emission from diffuse, optically thin gas at the
locations of some of the submm clumps. The contribution to the submm
emission from foreground molecular and atomic clouds using these CO
tracers was found to be negligible, but these emission lines may not
be tracing all of the gas in the ISM.

The three colour {\it Herschel} PACS image of Kepler is displayed in
Fig.~\ref{fig:3colkep}. The remnant is clearly seen in blue and
follows the same asymmetric structure as seen in the X-ray and radio
(\S\ref{sec:resultskepler}): the emission is brightest in the north
along an arc-like shell and some emission is seen towards the centre
of the remnant. The low level red structures seen across the map
(outside of the remnant) originate from dusty interstellar clouds with
temperatures $\sim 13-20\,\rm K$.

\begin{figure*}
   \centering 
   \includegraphics[trim=1mm 3mm 0cm
   0mm,clip=true,width=18cm]{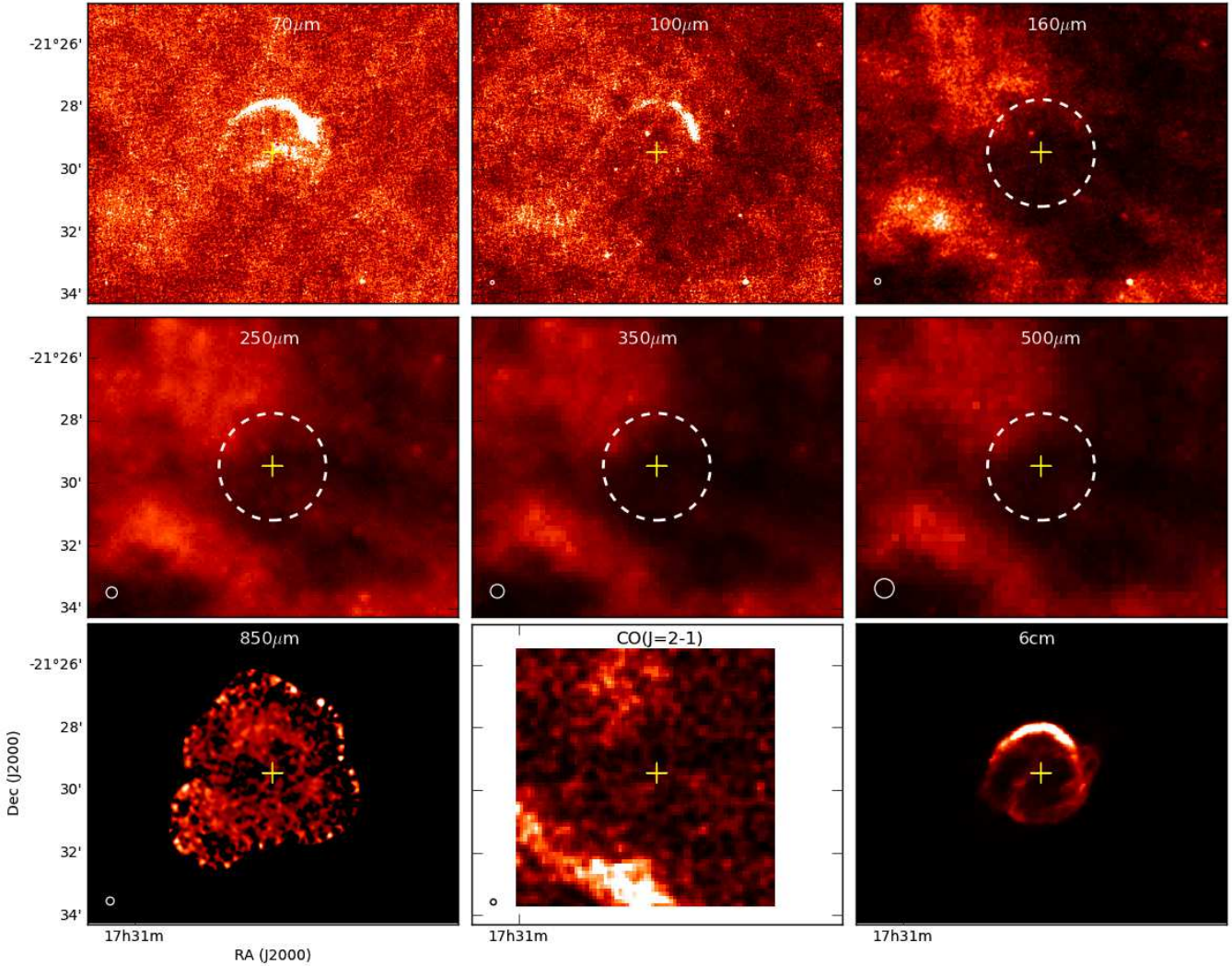}
   \caption{\small{Multiwavelength montage of Kepler's supernova remnant and
     its environs. {\it Top:} PACS and {\it middle:} SPIRE bands
     centred at $\rm RA=262.671^{\circ}$, $\rm Dec=-21.4914^{\circ}$
     (J2000.0; yellow cross); the region show is $7.2^{\prime} \times 7.2^{\prime}$.  The
     location of the forward shock (radius $103^{\prime \prime}$) is
     indicated with the white circle (DeLaney et al.\ 2002).  Also
     shown are {\it bottom}: the SCUBA 850\,$\mu$m (Morgan et al.\
     2003); the CO($J=2-1$) map integrated from $-200<v<200\,\rm km\,s^{-1}$ (Gomez et al.\ 2009) and the 6\,cm VLA
     map (DeLaney et al.\ 2002).}}
         \label{fig:kepmulti}
   \end{figure*}

\subsection{The multiwavelength view}
\label{sec:resultskepler}

\begin{figure*}
   \centering 
\includegraphics[trim=25mm 0mm 0mm 0mm,clip=true,width=18cm]{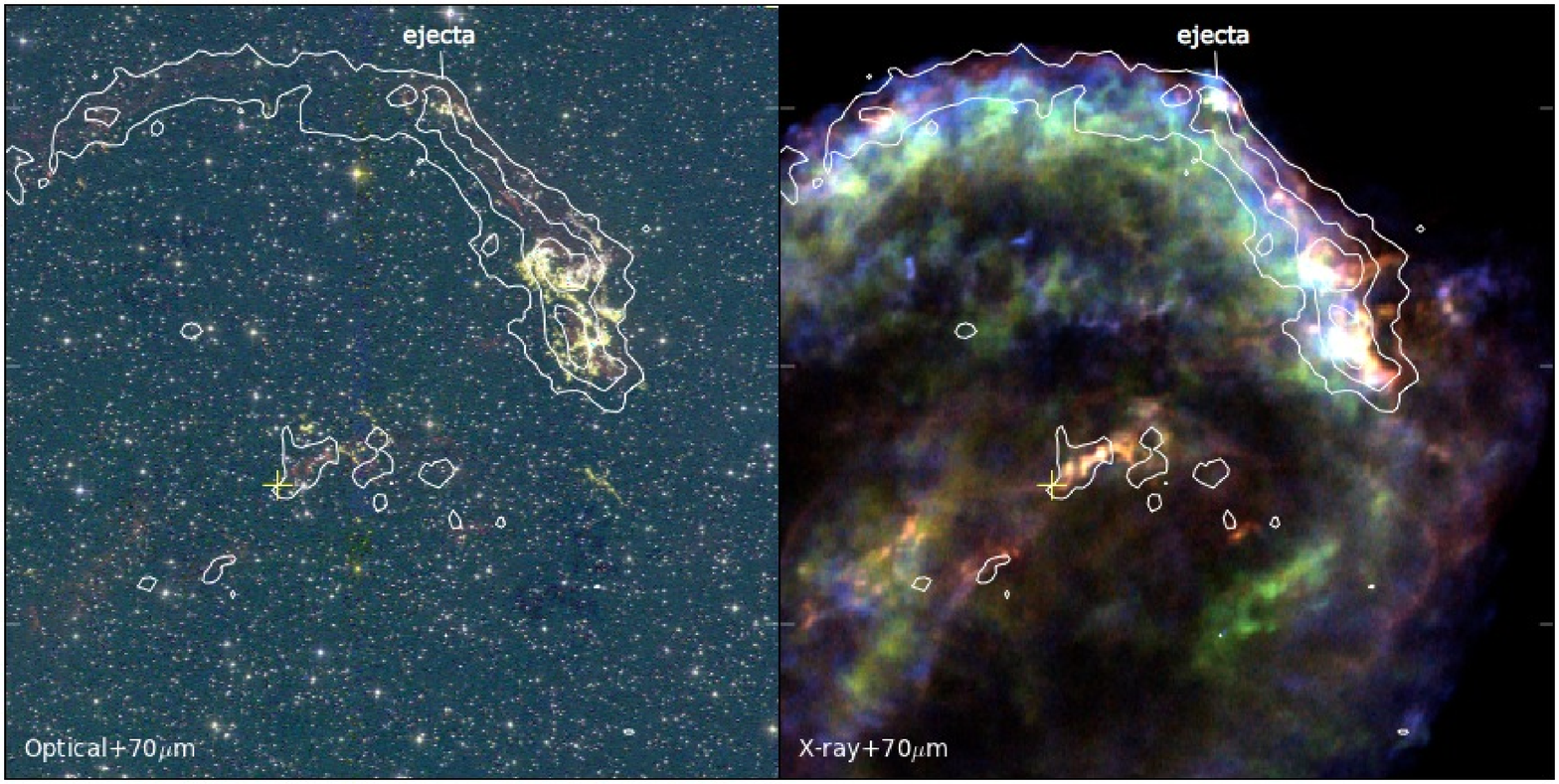}
\caption{\small{Comparison of the warm dust emission traced by PACS 70-$\mu$m
  data (white contours) with optical and X-ray.  {\it Left:} Three colour image of Kepler's supernova remnant
  with the {\it HST} at 658\,nm (red), 660\,nm
  (green) and 502\,nm (blue) indicating emission at $\rm H\alpha$ +
  [N{\sc ii}], [N{\sc ii}] and [O{\sc iii}] respectively.  The optical
  emission mostly arises from shocked circumstellar/interstellar
  material interacting with the primary blast wave.  The north
  component indicated with the fiducial is thought to be ejecta
  material (Sanrkit et al.\ 2008) which has overrun the primary shock
  boundary due to the contact discontinuity in the blast wave.  {\it Right:} Three colour
  X-ray image using {\it Chandra} archive data showing soft ($\rm
  0.3-0.72\,keV$: red), ($\rm 0.72-0.9\,keV$: green) and hard X-rays
  ($\rm 1.7-2.1\,keV$: blue).  The images are centered on $\alpha_{\rm
    J2000} = 262.663^{\circ}$ and $\delta_{\rm J2000} =
  -21.485^{\circ}$ and are $3^{\prime} \times 3^{\prime}$ across. }}
         \label{fig:kepxrayopt}
   \end{figure*}

   In Fig.~\ref{fig:kepmulti}, we present a multiwavelength view
   centred on Kepler's SNR, comparing the {\it Herschel} PACS and
   SPIRE photometric data with synchrotron emission from the remnant at
   6\,cm, unrelated molecular gas (traced by the CO emission) and
   850\,$\mu \rm m$ emission from cold dust. The complete shell-like
   structure of SN material is clearly seen at radio wavelengths,
   tracing the forward and reverse shocks (DeLaney et al. 2002).  In
   Fig.~\ref{fig:kepxrayopt} we compare the optical and X-ray emission
   with the warm dust from {\it Herschel}-PACS at $70\,\rm \mu m$.  The
   well-known north-south asymmetry seen at optical, radio and X-ray
   wavelengths with the enhanced northern emission (thought to
   originate from the motion of Kepler through the surrounding
   interstellar material) is also seen in the PACS data.  The `arm'
   structure across the middle (clearly seen in the radio, Fig.~\ref{fig:kepmulti}) originates
   from a projection effect of material in the front and back of the
   shell (Blair et al. 2007).  {\it Herschel} detects faint IR emission
   arising from this component at 70\,$\mu$m.  The faint
   `ear'-like features seen along the outermost midsection of the
   expanding shell at radio and X-ray wavelengths
   (Figs.~\ref{fig:kepmulti} \& \ref{fig:kepxrayopt}) are not seen in the
   infrared.  At the longer (160\,$\mu$m) PACS and the SPIRE
   wavelengths, it is difficult to disentangle any SNR emission from
   the large-scale interstellar structures extending across the
   region; the maps are increasingly dominated by unrelated dust
   clouds at temperatures of $\sim 16-20\,\rm K$.  The southern cloud
   of dust which is outside the area of the blast wave and ejecta
   regions is the same cloud orginally detected in molecular emission
   by Gomez et al. (2009) at 5\,$\rm km\,s^{-1}$. The cold dust cloud
   in the north-east is coincident with the CO structure at 10\,$\rm
   km\,s^{-1}$ (see \S\ref{sec:kepcold} for more discussion).

   Optical emission arising from dense, knotty regions in the
   circumstellar medium overrun by the blast wave, is traced in
   [NII] and $\rm H\alpha$.  Interestingly, the 70\,$\mu$m
   emission correlates with the $\rm H\alpha$ + [N{\sc ii}] features
   (Fig.~\ref{fig:kepxrayopt}) in both the northern limb (particularly
   in the north-west region) and across the projected shell along the
   middle.  Since the $\rm H\alpha$ emission originates from
   collisional heating of the thin region behind the SN shockfront
   (i.e. from swept-up surrounding material), the spatial agreement
   with the PACS emission and the ionised features suggest this dust
   component arises from the swept up medium.  This is further
   supported when comparing the FIR with X-ray emission.  There are
   three X-ray components in Kepler: continuum X-rays from the blast
   wave sweeping up surrounding circumstellar and interstellar
   material, X-ray line emission from the warm ejecta, and continuum
   emission arising from the non-thermal synchrotron emission in the
   north, south and eastern limbs.  The warm dust emission in the FIR
   is confined within the radius of the outer blast-wave and is seen
   only at the brightest points in the optical, radio and X-ray (Fig.~\ref{fig:kepxrayopt} - as
   noted in Blair et al.\ 2007) with peaks at 70\,$\mu$m seen where the
   soft and hard thermal X-rays overlap.  Therefore, the warm dust is
   associated with the densest gas behind the forward shock rather than the inner
    ejecta material; this confirms the dust originates from
   the outermost regions of the remnant i.e. from the swept up shocked
   circumstellar and/or interstellar medium (\S\ref{sec:origins}).

\subsection{The spectral energy distribution}
\label{sec:keplersed}

\begin{table*}
  \begin{tabular}{lcccccccc}\\ \hline \hline
 \multicolumn{1}{c}{}& \multicolumn{8}{c}{IR Fluxes Towards Kepler SNR (Jy)} \\
    &$24\rm \mu m$&$70\rm \mu m$ &$100\rm \mu m$& $160\rm \mu m$ & $250\rm \mu m$ & $350\rm \mu m$ & $500\rm \mu m$ & $850\rm \mu m$ \\ \hline 
   Published &  $9.5 \pm 1.0^a$ & $5.6 \pm 1.4^b$ & $2.9 \pm 1.1 ^b$& $<0.9^a$ & .. &.. &  $3.0 \pm 0.2^{c,d}$&$1.0\pm 0.2^c$ \\
   This work & $9.5 \pm 1.0^e$  & $12.3 \pm 2.7$ & $11.2\pm 2.4$ & $16.5\pm 2.9$ & $13.0 \pm 2.8$ & $5.8 \pm 1.2$ &$2.7\pm 0.6$ & $0.7 \pm 0.1$\\ 
    Nonthermal  & 0.02& 0.05&0.06 &0.09& 0.12&0.16 & 0.20& 0.29\\ \hline
\end{tabular}
\caption{\small{Fluxes in $\rm Jy$ measured for Kepler's remnant within the
  $120^{\prime \prime}$ aperture.
  Note that the previous published integrated fluxes use different
  aperture sizes to this work.  
  References: $^a$ - Blair et al.\ (2007);  $^b$ - average values from Douvion et
  al. (2001), Arendt (1989) and Saken, Fesen \& Shull (1992) {\it IRAS} data;
  $^c$ - Gomez et al.\ (2009); $^d$ - 450\,$\mu$m.
  $^e$ - Measured on {\it Spitzer} MIPS archive data. }}
\label{tab:fluxes}
\end{table*}

\begin{figure}
\includegraphics[trim=0mm 0mm 0mm
  0mm,clip=true,angle=-90,width=8.5cm]{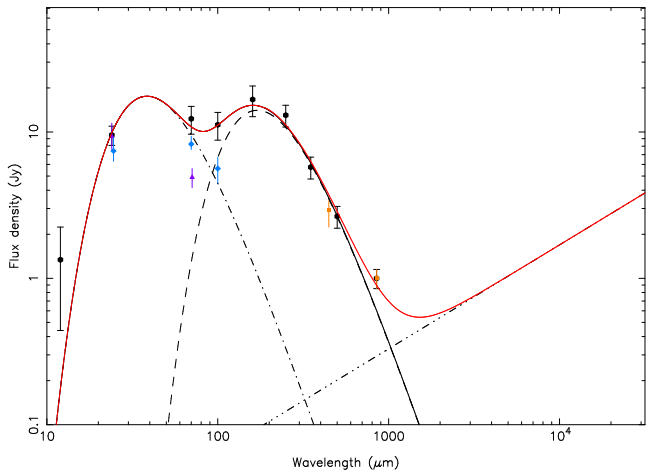}\hfill
\caption{\small{The IR-radio SED towards Kepler's SNR. The
    previously published fluxes from the literature
    (Table~\ref{tab:fluxes}) are shown with errors as vertical bars
    ({\it Spitzer} = purple triangles; {\it IRAS} = blue diamonds;
    SCUBA = orange squares).  The solid black line is the sum of the
    two-temperature thermal dust components fitted to the IR ($T_w =
    \rm 86$\,K and $T_c = \rm 20$\,K) and submm fluxes (dot-dashed
    and dashed lines respectively).  The dot-dot-dot-dashed line shows
    the non-thermal SED expected from the synchrotron, $S_{\nu}
    \propto \nu^{-0.71}$ scaled by the 6\,cm fluxes.  The red line
    shows the total SED from the non-thermal and thermal
    components. Note that the previous published integrated fluxes use
    different aperture sizes to this work. }}
\label{fig:sed} 
\end{figure}

The IR-submm spectral energy distribution (SED) towards Kepler's SNR
is shown in Fig.~\ref{fig:sed}. The fluxes estimated from {\it
  Spitzer}, {\it Herschel} and SCUBA in this work within an aperture
of radius $120^{\prime \prime}$ centred on the remnant, are plotted
with their errors (estimated from the calibration error, the error in
the zero level estimated in \S\ref{sec:zero} and from the variations
in the pixel RMS within the aperture).  All the data were smoothed to
the resolution of the SPIRE 500\,$\mu$m data using a Gaussian kernel.
The aperture is somewhat larger than the SNR boundary defined by the
edge of the radio ($103^{\prime \prime}$) but was chosen to encompass
the SNR emission in the images smoothed to the 500\,$\mu$m beam.  The
{\it Spitzer} fluxes were colour corrected using algorithms detailed
in the MIPS handbook\footnote{{\it
    http://irsa.ipac.caltech.edu/data/SPITZER/docs/mips/}}. We also
compare our fluxes with previously published fluxes in the literature
(Arendt 1989, Braun 1987, Douvion et al. 2001, Morgan et al. 2003,
Blair et al.\ 2007) although these were obtained using apertures with
radius ranging from $100-120^{\prime \prime}$ and therefore differ
from the fluxes in Table~\ref{tab:fluxes} accordingly. Note also that
we measure the flux encompassed in the aperture around Kepler, and therefore includes emission from
the foreground cloud.

The contribution to the IR/submm fluxes from synchrotron emission is obtained
by scaling the VLA radio map of Kepler using a constant spectral
index, $S_{\nu} \propto \nu^{\alpha}$ where the mean value of $\alpha$
is $-0.71$ but ranges from -0.6$-$-0.85 (DeLaney et al. 2002).  The synchrotron emission at
500\,$\mu$m is $0.20 \pm 0.06 \, \rm Jy$ ($\sim 15$\,per\,cent of the
total flux).  We
fit a two-component modified blackbody to the dust SED after the
synchrotron component has been subtracted; the model is the sum of two modified
Planck functions each with a characteristic temperature, $T_w$ and
$T_c$ (Eq.~\ref{eq:sed}):
\begin{equation} 
S_{\nu} =N_w\times \nu^{~\beta}B(\nu,T_w)+N_c \times \nu^{~\beta}B(\nu,T_c)
\label{eq:sed}
\end{equation}

where $N_w$ and $N_c$ represent the relative masses in the warm and
cold component, $B(\nu,T)$ is the Planck function and $\beta$ is the
dust emissivity index. The model was fitted to the SED (constrained by
the 12\,$\mu$m flux) and the resulting parameters ($N_w$, $N_c$,
$T_w$, $T_c$, $\beta$) which gave the minimal chi-squared were found.
We also applied a bootstrap analysis to our SED-fitting to determine
the errors on the SED model. The photometry measurements were
perturbed according to the errors at each wavelength, the new fluxes
were fitted in the same way, and the SED parameters recorded, with the
procedure repeated 1000 times.

\begin{figure}
\centering
\subfigure[]{\includegraphics[trim=0mm 0mm 6mm 0mm,clip=true,angle=0,width=9.2cm]{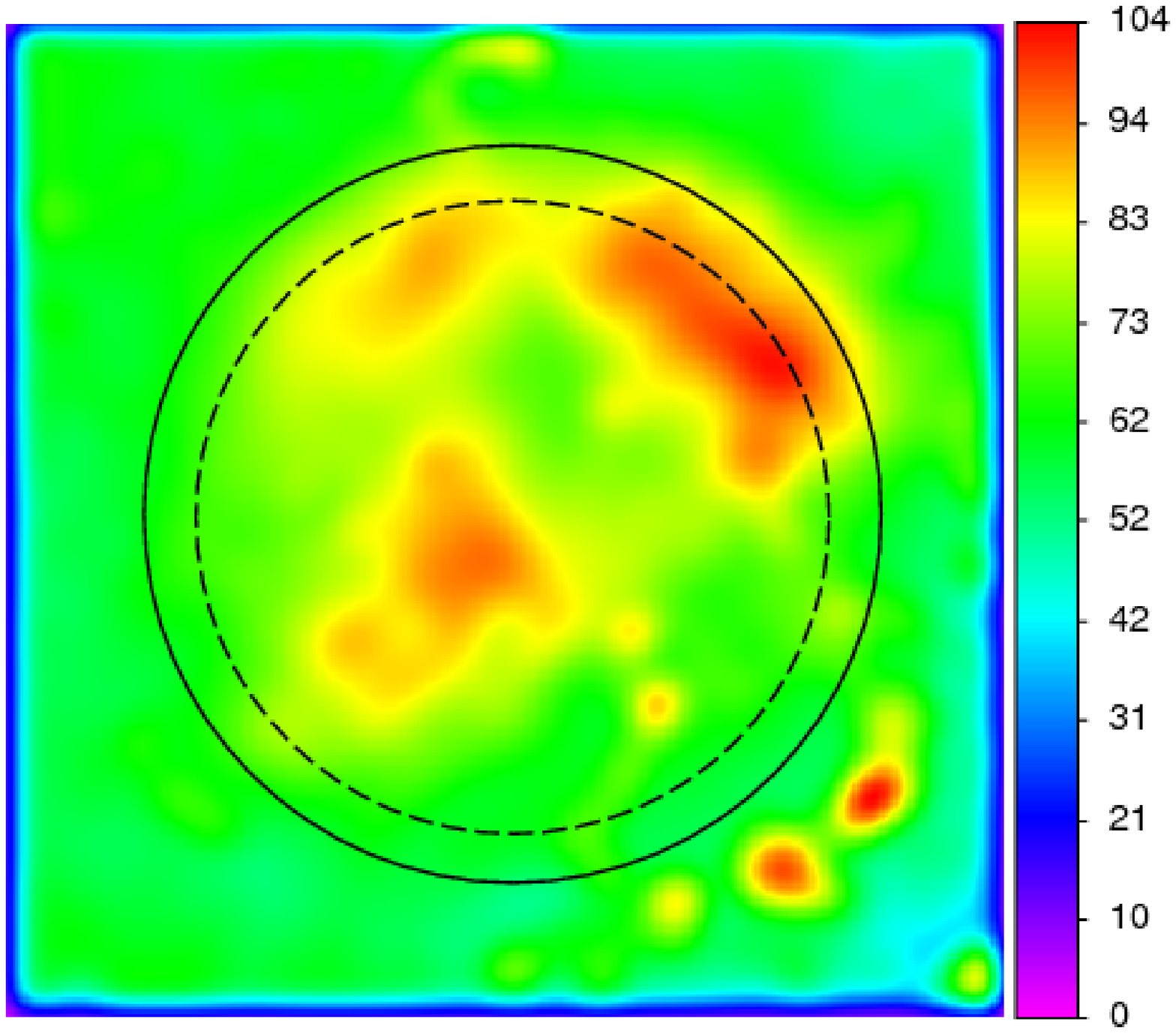}}\hfill
\subfigure[]{\includegraphics[trim=0mm 0mm 6mm
  0mm,clip=true,angle=0,width=9.4cm]{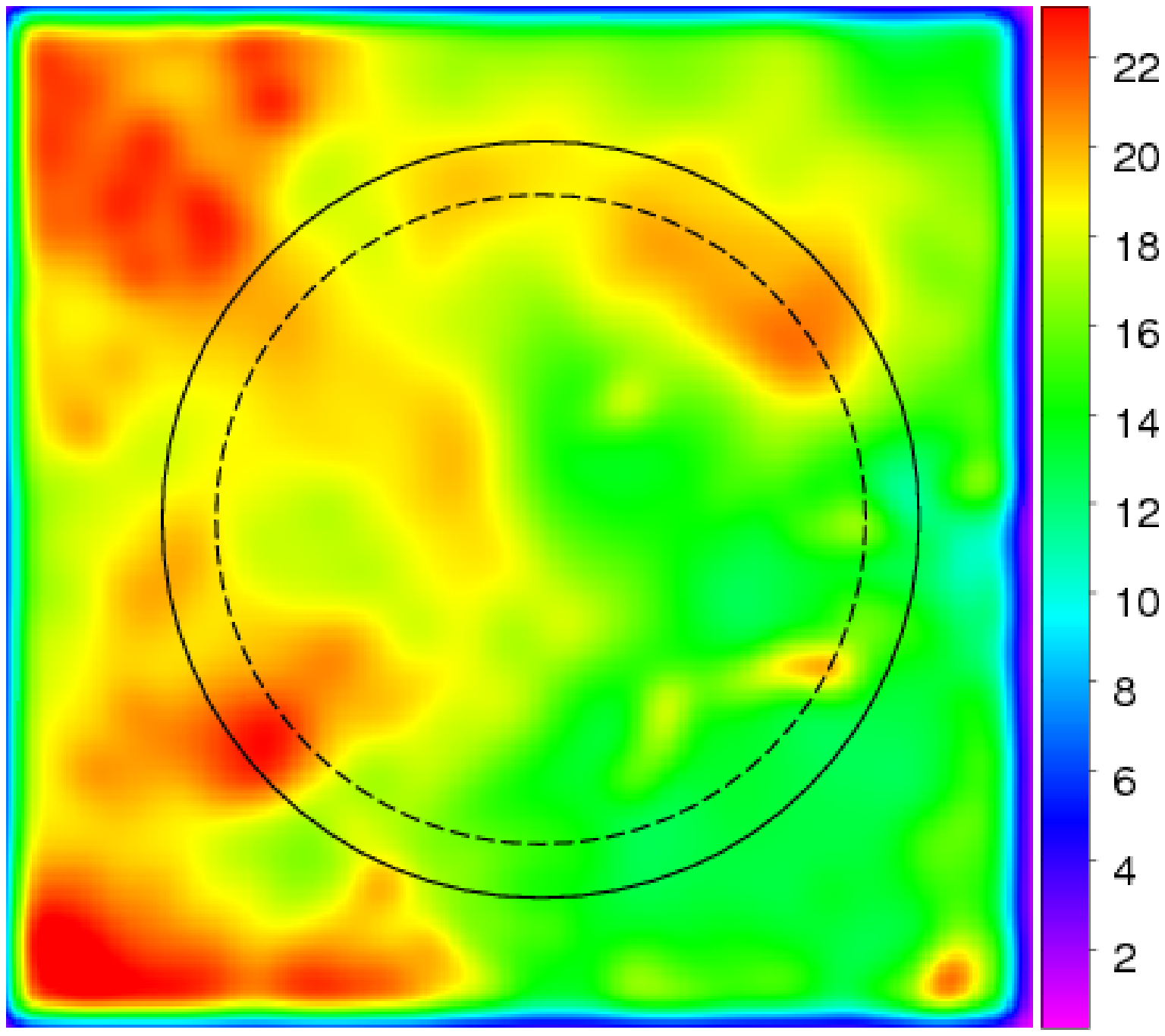}}\hfill
\caption{\small{Dust temperature maps derived from fitting a modified
    greybody to the 24-500\,$\mu$m
    maps towards Kepler's supernova
    remnant showing{\bf (a)} warm dust and {\bf (b)} cold dust.  The colourbar
    shows the temperature in K. The solid black circle indicates the
    photometry aperture used with radius $120^{\prime \prime}$ and the
    black dashed circle is the average radius of the outer shockfront
    $103^{\prime \prime}$. }}
\label{fig:keptemp} 
\end{figure}

The total dust mass is calculated using Eq.~\ref{eq:dustmass}:
\begin{equation}
  M_d = {S_{500} D^2 \over{\kappa_{500}}}\times \left[ {N_w \over{B({\rm 500}~\mu{\rm m}, T_w)}} +{ N_c\over{B({\rm 500}~\mu{\rm m}, T_c)}}\right]
\label{eq:dustmass}  
\end{equation}
where $\kappa_{500}$ is the dust absorption coefficient and $D$ is the
distance (we choose the 500\,$\mu$m flux since this is less sensitive
to the temperature). The distance to the remnant is uncertain, with
measurements of H{\sc i} absorption placing Kepler at $\sim 5\,\rm
kpc$ (see Reynoso \& Goss 1999 for a full discussion). Sankrit et al.\
(2008) have revised the measurement to $\rm 3.9^{+1.9}_{-1.4}\,kpc$
using HST proper motion data. We follow Blair et al.\ (2007) in
adopting a distance of $4\,\rm kpc$. To determine the dust mass, we
use $\kappa_{500}= 0.1\rm \,m^2\,kg^{-1}$ with $\kappa \propto
\lambda^{-\beta}$, appropriate for grains in the Milky Way and $\beta
\sim 1-2$ (Draine \& Lee 1984). The best-fit dust mass within the
aperture obtained from the SED model (Eqs.~\ref{eq:sed} \&
\ref{eq:dustmass}) is therefore $ M_w = 0.0035 \,\rm M_{\odot}$ and
$M_c = \rm 2.2 \,M_{\odot}$, with temperatures $T_w = 84\rm \,K$,
$T_c=\rm 19\,K$ and $\beta=1.5$. The temperature of the warm component
is consistent with collisional heating of dust in the shocked
gas with temperature and electron density of 1.2\,keV and $n_e
\sim 20\,\rm cm^{-3}$ (Bouchet et al.\ 2006, their Fig.~15).

The median parameters and confidence intervals from the bootstrap
method are listed in Table~\ref{tab:sed}. The dust mass depends
strongly on the choice of $\kappa$, for example, typical interstellar
dust grains are described well with $\kappa = 6.7\,\rm m^2\,kg^{-1}$
at 70\,$\mu$m (e.g. Draine \& Li 2001). In this case, the warm dust
mass would be reduced to $M_{w} =1 \times 10^{-3}\,\rm M_{\odot}$. We
note that the uncertainty in $\kappa$ could be an order of magnitude.

Temperature maps of the dust emission towards Kepler were created
using the convolved IR-submm maps from 24-500\,$\mu$m, and fitting a
two-temperature modified blackbody SED to each pixel. We extracted the
warm and cold components for each pixel, shown in
Fig.~\ref{fig:keptemp} (a) and (b). In order to account for the
non-thermal component, the flux due to synchrotron radiation was
subtracted from the dust maps on a pixel-by-pixel basis before the
greybody fits were made. The synchrotron contribution in each pixel
was estimated using a spectral index map created from the 6\,cm and
23\,cm VLA observations convolved to the same beam (see also DeLaney
et al.\ 2002). The somewhat flatter spectral index in the northern
bright ring ($\alpha \sim -0.68$ compared to the average value of
-0.71) resulted in removing most of the cold dust in this region,
though there is residual emission associated with this component
within the SNR shell (Fig.~\ref{fig:keptemp} (b)). This cold dust
feature appears to be coincident with a similar structure seen in the
warm dust component but the cold dust is further out from the centre
by $\sim 10^{\prime \prime}$ compared to the warm dust emission which
is spatially coincident with the radio (Fig.~\ref{fig:keptemp} (a)).

The cold dust structures in the north are also coincident with the
structures seen at 850\,$\mu$m with the SCUBA camera (Gomez et al.\
2009, Fig.~\ref{fig:kepmulti}) originally attributed to supernova dust
due to the lack of correspondance between the CO gas from molecular
interstellar clouds towards Kepler (Fig.~\ref{fig:kepmulti}) and the
850\,$\mu$m emission. The larger {\it Herschel} map clearly shows large-scale
and exended interstellar dust structures right across the entire
region towards Kepler and implies that that the argument in Gomez et
al.\ was flawed.  This could simply be due to a non-constant ratio of
   molecular gas-to-dust, or that the $^{12}$CO and $^{13}$CO
   lines are not effective tracers of the gas in this region. Alternatively, it is
   possible that, towards Kepler, the interstellar dust structures are
   associated with atomic rather than molecular gas.

   If the cold dust in the north of the remnant is located at the distance of the
   remnant, the location of the emission
   is suggestive of dust swept up by the blast wave, though the mass
   of dust in the northern shell is rather substantial at $\sim 1.0\rm
   \,M_{\odot}$ for $T\sim 20\,\rm K$. Elsewhere, the cold dust
   component does not spatially coincide with any of the SNR tracers
   including X-ray, radio and optical emission and is unlikely to be
   related to the SN itself. We discuss the origin of the cold dust
   further in \S\ref{sec:kepcold}.

Finally, we also tested for the presence of cool dust with temperatures in the range
30-40\,K as found in Cas A (Sibthorpe et al.\ 2010; Barlow et al.\ 2010): the flux
contribution of the warm component (visible at 24\,$\mu$m) was
subtracted from the 70 and 100\,$\mu$m maps, however, we find no
evidence of a cool dust component.

\section{Tycho's Supernova Remnant}
\label{sec:tycho} 

The Tycho supernova event was observed by Tycho Brahe in 1572. It lies at a
distance of $1.5-3.8$\,kpc (Reynoso et al.\ 1997; Krause et al.\
2008b) and has a thin shell-like morphology with diameter $\sim
8^{\prime}$ across.  Here, we use the distance given in $D=3.8\,\rm kpc$ as this is consistent with the historical records of the explosion, the non-detection of the remnant and the distance to the proposed binary companion (Lu et al. 2011) (see Krause et al.\  2008b for more details).  The outer blast wave extends to an average distance of $251^{\prime
  \prime}$ from the centre, with the contact discontinuity between the
reverse shock material and swept up ISM at angular distances of $183$,
$241$ and $260^{\prime \prime}$ depending on the position angle
(Warren et al.\ 2005). Tycho's light echo spectrum has revealed it to
have been a Type Ia event (Krause et al. 2008b); it is the remnant of
a canonical single-degenerate binary explosion (Lu et al. 2011) as
evident from X-ray ejecta abundances (Decourchelle et al.\ 2001). This
is further supported by the lack of similarity in Tycho's optical
light curve with typical sub-luminous (van den Bergh 1993) and
over-luminous (Ruiz-Lapuente 2004) Type Ia explosions (Krause et al.\
2008b). Badenes et al.\ (2006) performed detailed comparison between
the X-ray spectra and the results from hydrodynamical modeling of the
supernova shocks, and found that the delayed detonation explosion
model was the best fit to the data. In this scenario, the SNR is
expanding into an ambient density of $\sim 0.6-3\,\rm cm^{-3}$, with
an ejecta mass of $\rm 1.3\,M_{\odot}$.

Many authors have suggested that Tycho is interacting with molecular
clouds (Reynoso \& Goss 1999; Lee, Koo \& Tatenatsu 2004; Cai, Yang \&
Lu 2009; Xu, Wang \& Miller 2011) particularly with a large CO
structure seen in the north and north-east at velocities between
-63.5 and -61.5$\,\rm km\,s^{-1}$; this structure is believed to be
responsible for the asymmetry seen in the X-ray and radio (see
Fig.~\ref{fig:multitycho}).  Cai et al. (2009) showed that interaction
with the ISM occurs along the entire boundary of the remnant with
interstellar structures detected at velocities -69 $-$ -58$\,\rm
km\,s^{-1}$; the total gas mass of clouds interacting with Tycho is
$10^3\,\rm M_{\odot}$ (Xu et al.\ 2010). Conversley, Tian \& Leahy
(2011) propose that the CO clouds are not interacting with the remnant
since the densities of the molecular gas clouds in the north are $\sim 200\,\rm
cm^{-3}$, compared to the X-ray ejecta parameters which suggest the remnant is
ploughing into material of density $0.2-3\,\rm cm^{-3}$ (Decourchelle
et al.\ 2001, Badenes et al.\ 2006). The difference in the density
estimated for the surrounding material from CO/X-ray abundances is a
factor of 60 (rather than the three orders of magnitude quoted in Tian
\& Leahy (2011)). However, the density of the H{\sc i} cloud they detect at 
-43 $-$ 57$\,\rm km\,s^{-1}$ is comparable to the density obtained from X-ray
observations and could therefore be tracing gas interacting with the
remnant.  Using a flat galactic rotation model (with standard
constants) places this H{\sc i} cloud at $d=3.4\,\rm kpc$ which is
consistent with the distance of Tycho's SNR.

Ishihara et al.\ (2010) detect warm dust features in the north of the remnant
with {\it AKARI} and suggest these are linked to the CO structures in the
northeast. They also find enhanced dust emission in the northwest
region of the shell and since they find no corresponding CO feature,
they suggest this component could originate from SN dust. Without the
long wavelength data, they were unable to constrain the amount of
cold dust arising from SN material within the blast wave.

\begin{figure}
   \centering 
\includegraphics[trim=0mm 0mm 100mm
   1mm,clip=true,width=8.5cm]{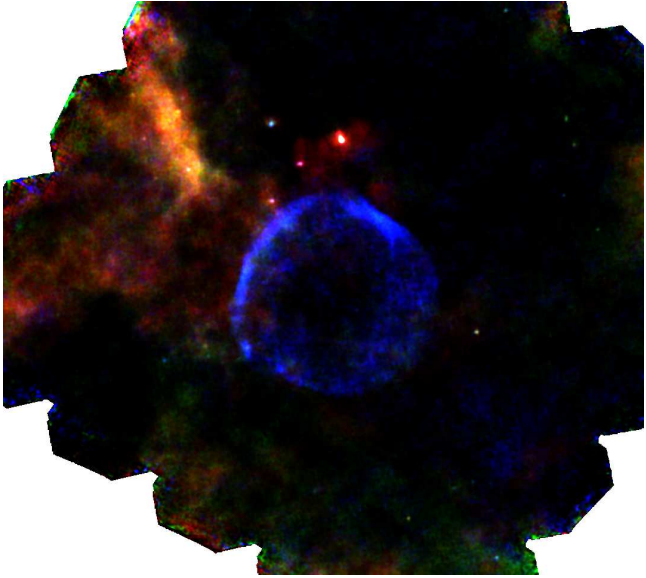}
   \caption{\small{Three-colour FIR image of Tycho's SNR with PACS data at
     70, 100 and 160\,$\mu$m. The area shown is $23.2^{\prime} \times
     27.2^{\prime}$. }     }
   \label{fig:3coltycho}
   \end{figure}

The three colour {\it Herschel} PACS image is presented in
Fig.~\ref{fig:3coltycho}. As in Kepler, the remnant is clearly visible in blue,
and the interstellar structures become clearer at the longer PACS and
SPIRE wavelengths (Fig.~\ref{fig:multitycho}). Cold, dusty clouds
peaking at 250\,$\mu$m are seen to the immediate east and west of the
remnant and have temperatures typical of the ISM.  The bright SPIRE
interstellar structure to the immediate north of the remnant in Fig.~\ref{fig:multitycho}, is
highly suggestive of emission originating from gas and dust swept up
by the expanding SNR. The structure is similar in appearance to the
submm cavities blown out by the winds from warm OB stars
(e.g. Schneider et al.\ 2010) with filamentary fingers pointing
inwards to the centre of the ionising radiation (in this case the
SNR). In the higher resolution PACS data (Fig.~\ref{fig:3coltycho}),
we see three small, warm sources embedded within the SPIRE clouds (see
\S\ref{sec:tychosed} for further discussion). The compact blue source
to the north (Fig.~\ref{fig:3coltycho}) is the Cepheid
Variable star, $\rm V^*$ AS Cas.

\begin{figure*}
   \centering 
\includegraphics[trim=0mm 10mm 0mm
   0mm,clip=true,width=18cm]{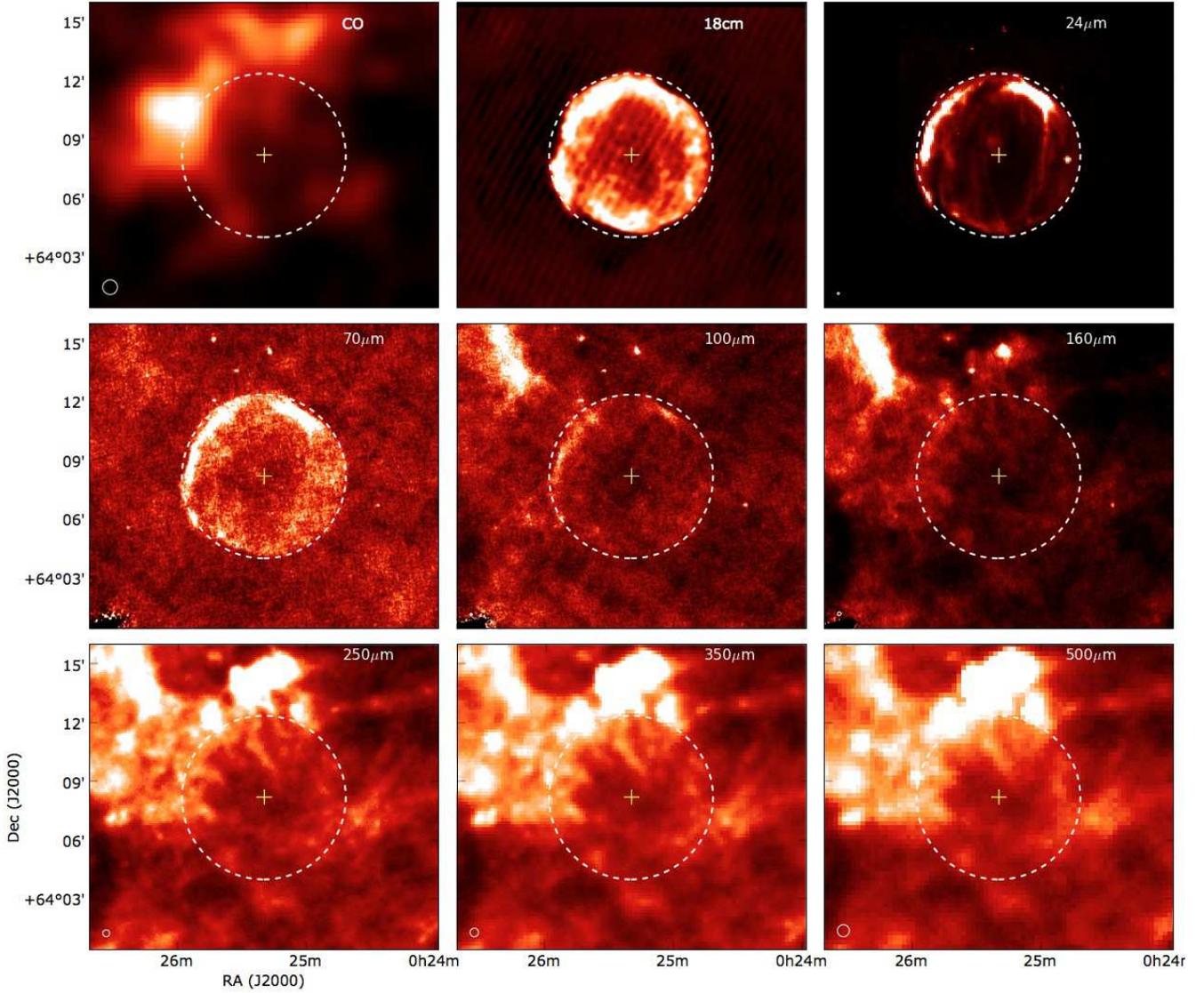}
   \caption{\small{Multiwavelength montage of Tycho's supernova remnant. From
     left to right: {\it top:} $^{12}$CO($J=1-0$) data integrated from
     $-68 <v<-53\,\rm km\,s^{-1}$), 18\,cm radio data and {\it
       Spitzer} MIPS 24\,$\mu$m. {\it Middle:} {\it Herschel} PACS and
     {\it bottom:} SPIRE.  The images are centred at $\rm RA=
     6.3308^{\circ}$, $\rm Dec= 64.1372^{\circ}$ (J2000.0; yellow cross)
     with size $9^{\prime} \times 9^{\prime}$.  The location of the forward
     shock (radius $251^{\prime \prime}$) is indicated with the white
     circle.}}
         \label{fig:multitycho}
   \end{figure*}

\subsection{The multiwavelength view}
\label{sec:resultstycho}

\begin{figure}
   \centering 
   \includegraphics[trim=25mm 5mm 0mm
   0mm,clip=true,width=8.5cm]{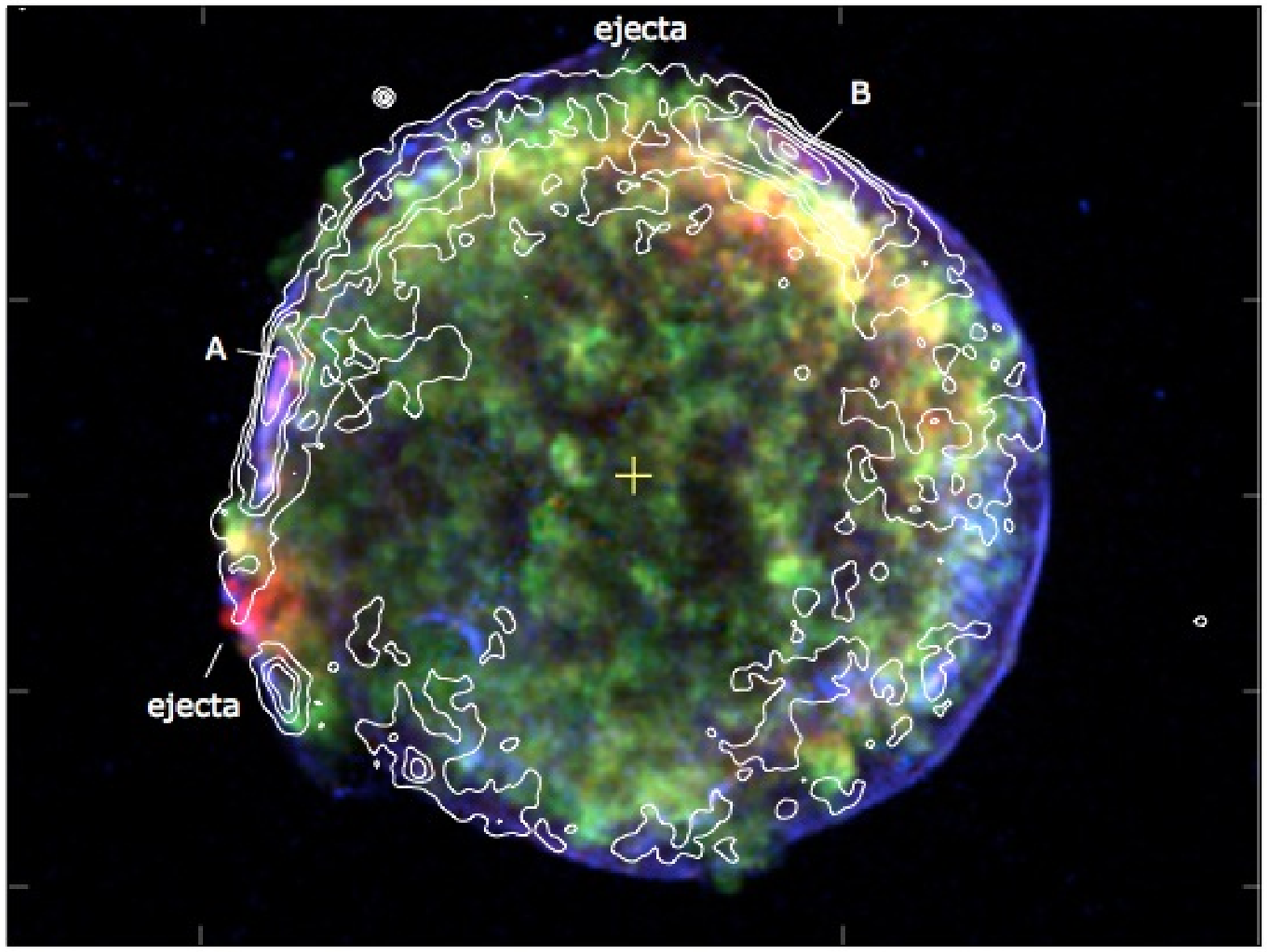}
   \caption{\small{Three colour X-ray image of Tycho's SNR using {\it
       Chandra} archive data showing soft thermal X-rays from ejecta
     and continuum ($\rm 0.95-1.26\,keV$: red) and hard X-rays tracing
     the both thermal and non-thermal continuum components ($\rm
     4.1-6.1\,keV$: blue).  The green X-rays trace the ejecta emission
     ($\rm 1.63-2.26\,keV$). White contours are 70\,$\mu$m PACS
     emission ($7, 9, 10, 13\, \rm mJy/pixel$).  Two regions where the
     ejecta is protruding out of the reverse shock out to the edge of
     the forward shock are labeled.  Labels `A' and `B' indicate regions where FIR emission is coincident with soft and hard X-rays.}}
         \label{fig:tychoxray}
   \end{figure}

\begin{figure}
   \centering 
   \includegraphics[trim=20mm 5mm 0mm
   0mm,clip=true,width=10cm]{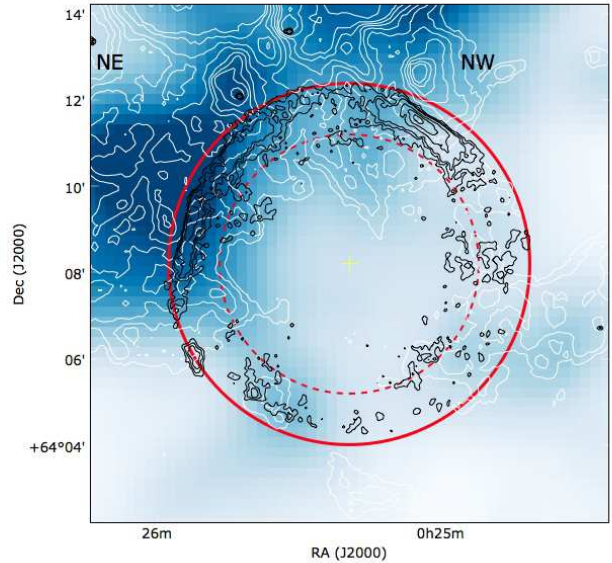}
   \caption{\small{Comparison of the warm and cold dust with {\it Herschel}
     (contours) and molecular gas (inverted colourscale). Black
     contours trace 70\,$\mu$m PACS emission from warm dust ($0.4, 5,
     6, 7, 9, 10, 13, 16, 20\, \rm mJy/pixel$). White contours trace
     the cold dust emitting at 250\,$\mu$m ($20, 50, 100, 150, 200,
     250, 300, 400\, \rm mJy/beam$). The red solid circle traces the
    forward shock wave at radius $251^{\prime \prime}$, the
     red dashed circle traces the reverse shock boundary at radius
     $183^{\prime \prime}$ (Warren et al.\ 2005). The forward shock is
     at radius $251^{\prime \prime}$. Following Ishihara et al.\
     (2010), the major interstellar CO clouds to the north of the
     remnant are labeled NW and NE. Note that the peaks in the
     250\,$\mu \rm m$ emission from cold dust are spatially coincident
     with the unrelated molecular material and not the warm dust in
     the SNR.  The contribution to the 250\,$\mu$m emission from
     synchrotron emission is $<6$\,per\,cent. }}
         \label{fig:tychoco}
   \end{figure}

The emission from warm dust at 70\,$\mu$m in the north has two bright
components in the NE and NW (Fig.~\ref{fig:multitycho}); seen also in
the {\it AKARI} data at wavelengths out to $160\,\mu$m. This
corresponds spatially to peaks in 24\,$\mu$m emission as well as the
X-ray continuum which traces the forward blast wave. The longer
wavelength {\it Herschel}-SPIRE data do not spatially correlate with
radio or X-ray emission from the remnant. In Fig.~\ref{fig:tychoxray},
we compare the PACS 70\,$\mu$m emission with a three-colour X-ray
image (0.95-1.26, 1.63-2.26 and 4.1-6.2\,keV). The emission from warm
dust at 70\,$\mu$m is bounded within the forward and reverse shocks of
the SNR at $251^{\prime \prime}$ and $183^{\prime \prime}$
respectively (Warren et al.\ 2005) and we do not see FIR emission
towards the interior of the remnant. The 250\,$\mu$m emission
(Fig.~\ref{fig:multitycho}), however, extends far outside the shock
boundary and spatially correlates with the low-level structures and
the peaks seen in the CO map (Fig.~\ref{fig:tychoco}).  There doesn't
appear to be any agreement between the peaks at 70 and 250\,$\mu$m
though we see emission at both FIR wavelengths where there is also CO
emission. Comparing with the CO molecular gas, the interstellar clouds
seen by {\it Herschel} close to the remnant in the north (labeled NW
and NE) and to the east, match structures seen at $-61$ and $-65\,\rm
km\,s^{-1}$. In the southeast and southwest, two faint CO clouds are
seen in the same range of velocity, both with associated 70\,$\mu$m
and 250\,$\mu$m emission.

\begin{figure*}
   \centering 
   \includegraphics[trim=35mm 10mm 0mm
   0mm,clip=true,width=17.5cm]{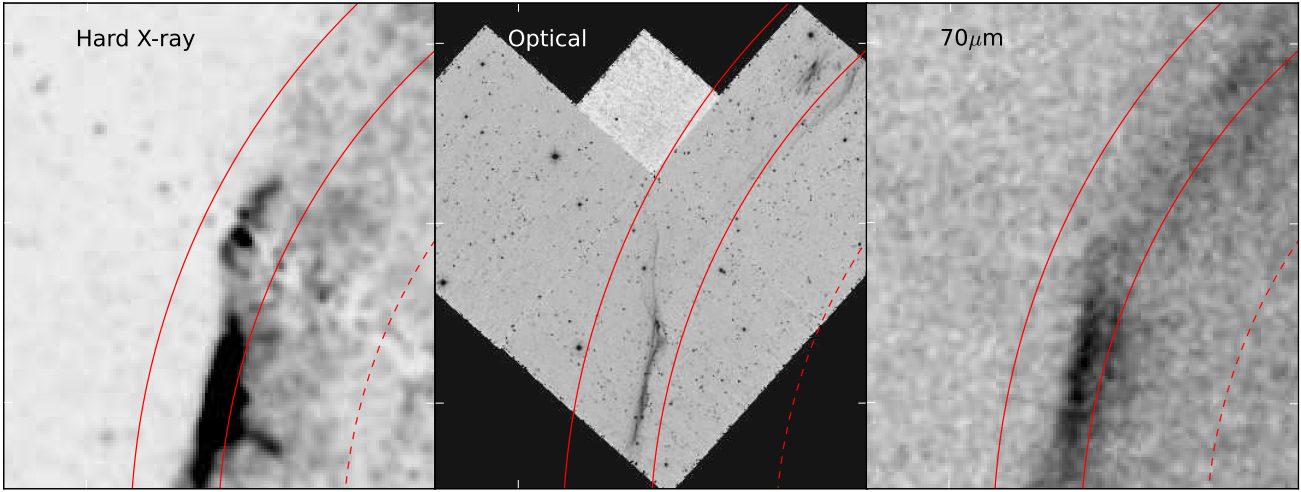}
   \caption{\small{ Zoomed in region on the eastern limb of Tycho's remnant:
     {\it left:} hard X-Ray {\it middle:} H$\alpha$ emission and {\it
       right:} 70\,$\mu$m emission.  The blastwave is at
     radius $251^{\prime \prime}$ and extends inwards to $222^{\prime
       \prime}$ (enclosed by solid red circles). The reverse shock at
     $183^{\prime \prime}$ is indicated by the red dashed line.  In
     general, the hard X-ray emission is a combination of non-thermal and
     thermal emission from the interaction of the supernova and the
     surrounding material.  The spatial agreement with the $\rm
     H\alpha$ emission suggests that the hard X-ray emission is dominated by thermal
     radiation in this region.  The image is centred at $\rm RA =
     6.4718^{\circ}$ and $\rm Dec = 64.1372^{\circ}$ with radius
     $72^{\prime} \times 72^{\prime \prime}$. }}
         \label{fig:tychoopt}
   \end{figure*}

The correlation between the 70\,$\mu$m emission and the X-rays emitted at
the outer edges of the SNR is striking. The hard X-ray emission
(4.1-6.2\,keV) is filamentary in structure with thin rims extending
outside of the circular edge both in the north and south-east side of
the remnant. The peaks in the warm dust trace the peaks in the hard
X-rays. The FIR emission does not spatially coincide with the soft
X-rays (Fig.~\ref{fig:tychoxray}) with peaks in the soft X-ray lagging
behind the 70\,$\mu$m peaks by $27^{\prime \prime}$ in the northwest,
though the soft emission at the edges of the outer blast wave do
correlate with the FIR (regions `A' and `B' in Fig.~\ref{fig:tychoxray}). The peaks in the radio emission also lag by
$\sim 16^{\prime \prime}$.  The ejecta material does not spatially
correlate with the IR emission, with the peak in the northwest lagging
behind the FIR by $23^{\prime \prime}$. The correlation of warm dust
with the hard X-ray emission at the edges of the remnant is slightly
surprising due to the large fraction of non-thermal emission expected
from shocks in this energy regime. However, dust is seen where soft
X-rays (from thermal emission due to the blast wave sweeping up
interstellar material) and hard X-rays overlap. Lee et al.\ (2010)
showed that the hard X-rays also contain a significant fraction of
thermal emission, we therefore conclude that the warm dust is spatially
correlated with the swept up material. The lack of 70\,$\mu$m emission
in the SW region of the remnant, where the hard X-rays are bright, may
suggest that the CO (Fig.~\ref{fig:tychoco}) and 250-\,$\mu$m
structure in this region orginates from a foreground cloud which is
not associated with the remnant; this would mean very little
interstellar dust is being swept up on this side (in agreement with
the faster expansion rate of the SNR observed in this region). The
hard X-rays in the west-southwest region are produced by non-thermal
synchrotron emission revealed by the presence of X-ray `stripes' due
to tangled magentic fields on small scales (Eriksen et al.\
2011). Here then the hard X-rays are dominated by non-thermal
processes and the lack of warm dust in this region further supports the
hypothesis that we only see dust in the shocked, swept-up ISM where we
find thermal continuum X-rays and $\rm H\alpha$ emission - this is
highlighted in Fig.~\ref{fig:tychoopt}.

There is no correlation between the FIR and X-rays for the outermost ejecta
material which has broken through the reverse shock region out to the
very edge of the SNR (as labeled in Fig.~\ref{fig:tychoxray}). We
conclude that there is no evidence to suggest that the warm dust is
associated with the ejecta.

\subsection{The spectral energy distribution}
\label{sec:tychosed}

\begin{table*}
  \begin{tabular}{lccccccccc}\\ \hline \hline
    \multicolumn{1}{c}{}& \multicolumn{9}{c}{IR Fluxes towards Tycho SNR (Jy)} \\
    &$12\rm \mu m$&$24\rm \mu m$&$70\rm \mu m$&$100\rm \mu m$& $140\rm \mu m$& $160\rm \mu m$ & $250\rm \mu m$ & $350\rm \mu m$ & $500\rm \mu m$  \\ \hline \hline
    Published &  $1.8 \pm 0.8^a$ & ..& $38.6 \pm 7.7^{b,c}$ & ..& $46.7\pm 14.0^{b}$ & $44.5\pm 13.4^{b}$ & ..& .. &..\\
 & .. &$28.3 \pm 3.2^{b,d}$  & .. &$40.6 \pm
    8.1^{b,e}$  &.. &..  &.. & ..& ..\\
    This work &  .. & $20.1\pm 3.0$ & $44.8\pm 10.8$& $41.1\pm
    8.8$ &  ..& $59.0\pm 14.1$
    & $42.7\pm 8.6$  & $32.8\pm 6.5$  & $17.6\pm 3.5$\\  
&  .. & $18.9\pm 3.8^f$ & & &  & 
    &  &  & \\  
Nonthermal  & ..& 0.65 & 1.28 & 1.58& 1.98&2.09 &2.73 &3.34 &4.14 \\ \hline
\end{tabular}
\caption{\small{Fluxes in $\rm Jy$ measured towards Tycho's remnant within the
  $251^{\prime \prime}$ aperture defined in Fig.~\ref{fig:multitycho}.
  References $^a$- Douvion et al.\ 2001; $^b$ - Ishihara et al.\ 2010;
  {\it AKARI} fluxes measured at wavelengths: $^{c}$ - $65\mu$m;
  $^{d}$ - $22.9\mu$m; $^{e}$ - $90\mu$m. $^f$ Measured on WISE data
  at 22\,$\mu$m.} }
\label{tab:tychofluxes}
\end{table*}

\begin{figure}
\includegraphics[angle=-90,width=8.5cm]{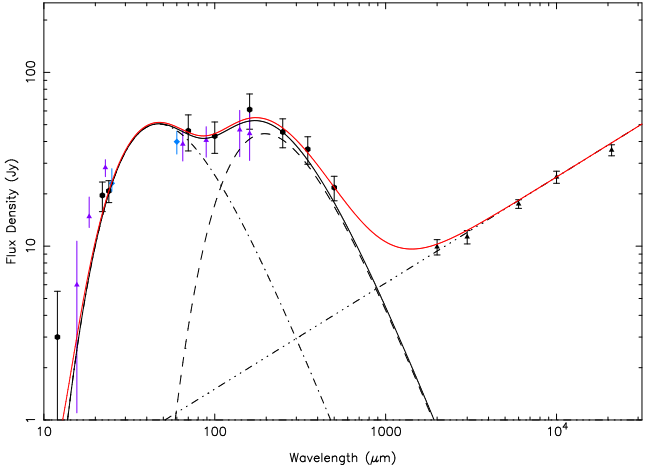}\hfill
\caption{\small{ The IR-radio SED of Tycho's SNR. The previously
    published fluxes from the literature (Table~\ref{tab:fluxes}) are
    shown with errors as vertical bars (IRAS = blue diamonds; WISE =
    blue triangles; {\it AKARI} = purple triangles from Ishihara et
    al.\ (2010)). (Note that the latter study used an aperture of
    diameter $525^{\prime \prime}$ compared to the $504^{\prime
      \prime}$ in this work.)  The solid black curve is the sum of the
    two-temperature thermal dust components fitted to the IR and submm
    fluxes (dot-dashed and dashed lines respectively) at temperatures
    $T_w =90\,\rm K$ and $T_c =21\,\rm K$.  The red curve shows the
    total SED from the non-thermal and thermal components. The
    synchrotron power law fit to the radio data is shown by the
    dot-dot-dot-dashed line.}}
\label{fig:tychosed} 
\end{figure}

The IR-submm SED towards Tycho's SNR is
shown in Fig.~\ref{fig:tychosed}. The fluxes estimated from {\it
  Spitzer} and {\it Herschel} within an aperture of radius
$251^{\prime \prime}$ centred on the remnant, are plotted with their
errors (Table~\ref{tab:tychofluxes}; the zero-level offsets were
applied to the maps before fluxes were measured - \S\ref{sec:zero}).
As the synchrotron emission in Tycho has been steadily decreasing with
a decay rate of $0.5~\% \rm~ yr^{-1}$ (Klein et al. 1979), the radio
fluxes were corrected for this decay rate to represent the same epoch
as the {\it
  Herschel} observations.  The radio spectral slope exponent in the
literature ranges from $\alpha = -0.52$ (Katz-Stone et al. 2000) to
$-0.61$ (Green 2001); for $\alpha=-0.61$ the contribution of synchrotron
to the 500\,$\mu$m flux is 4\,Jy (17\%), for $\alpha=-0.52$ this
becomes 5\,Jy.  We use the former spectral index in this work since
this describes the observed radio fluxes (Fig.~\ref{fig:tychosed}).
The submm flux within the aperture defined by the outer
blast wave is therefore well above the expected synchrotron emission.

\begin{figure}
\centering
\subfigure[]{\includegraphics[trim=0mm 0mm 18mm 0mm,clip=true,angle=0,width=8.5cm]{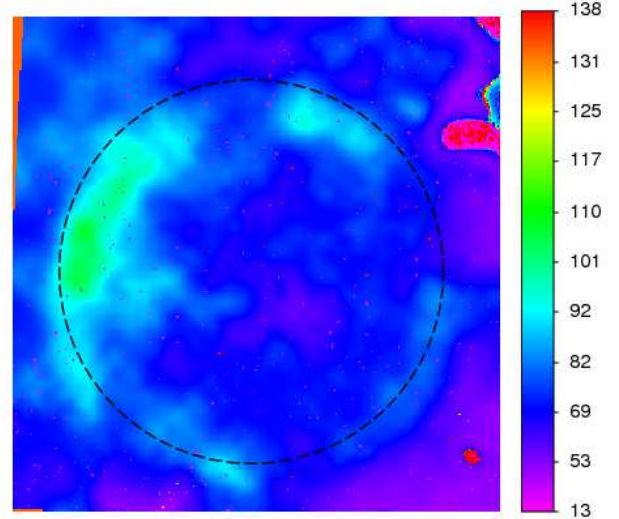}}\hfill
\subfigure[]{\includegraphics[trim=0mm 0mm 18mm
  0mm,clip=true,angle=0,width=8.5cm]{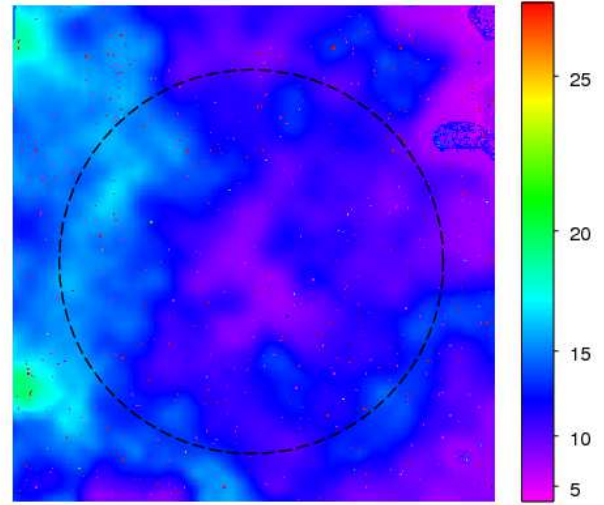}}\hfill
\caption{\small{Dust temperature maps derived from fitting a modified
    greybody to the 24-500\,$\mu$m maps for the area surrounding
    Tycho's supernova remnant with SED components separated into {\bf (a)} warm dust and {\bf (b)}
    cold dust. The colourbar shows the
    temperature in $\rm K$. The solid black circle indicates the
    forward shock radius at $251^{\prime
      \prime}$. }} \label{fig:tychotemp}
\end{figure}

As described in \S\ref{sec:keplersed}, we fit the IR-radio SED with
two-temperature modified blackbodies and a power law spectrum. The
best fit parameters from the SED are $T_w = 90\rm \,K$, $T_c=\rm
21\,K$, $\beta=1.5$ and dust mass $M_w =8.6\times 10^{-3}$ and $M_c= 4.0
\rm \,M_{\odot}$.  The median parameters and confidence intervals from
the bootstrap method are listed in Table~\ref{tab:sed}. The
temperature maps are shown in Fig.~\ref{fig:tychotemp}.  The warm dust
is clearly seen within the remnant at the same location as the
24\,$\mu$m emission and the outermost radio emission.  The cold dust
map highlights a dust structure on the eastern edge which appears to
overlap with the warm dust (Fig.~\ref{fig:tychotemp} (b)) and appears
to be swept up material.  If this structure is at the distance of
Tycho with $T_d \sim 21\, \rm K$ and absorption properties similar to
interstellar dust, the dust mass would be $\rm \sim 2\,M_{\odot}$,
well above the mass of heavy elements available to form dust.

As with Kepler, we find no evidence for a separate cool dust component
(at $T_d \sim \rm 30-40\,K$) in the SNR when subtracting the hot dust
traced by emission at 24\,$\mu$m from the 70 and 100\,$\mu$m maps.

\subsection{An aside on Triggered Star Formation}
Returning to the three FIR-bright, small clumps seen in
Fig.~\ref{fig:multitycho}, one may jump to the conclusion that the
remnant has triggered star formation in the compression of material as
the blast wave moves outwards. Triggered star formation has long been
suggested as a viable star forming mechanism (e.g. Elmegreen \& Lada
1977), where an expanding H{\sc ii} region compresses gas between the
ionization front and the shock front, causing fragmentation into
gravitationally unstable, star-forming cores. Supernova-triggered star
formation has been inferred through evidence of supernova-shocked
molecular gas (e.g. Reach \& Rho 1999, Reynoso \& Magnum 2001). The
FIR-bright clumps seen by {\it Herschel} are detected at 24\,$\mu$m,
suggesting these are star-forming cores rather than protostars. The
temperatures we obtain from fitting their SEDs from 24-500\,$\mu$m
suggest each core has a warm and cold dust temperature component with
$53-57\rm \,K$ and $\sim 12\rm \,K$ with dust masses ranging from $\rm
1.4-2.8\,M_{\odot}$ for the two smaller cores and $14\,\rm M_{\odot}$
for the largest and brightest core. These masses were estimated using
standard interstellar $\kappa$ values and (assuming they are at the
distance of Tycho) indicate the formation of intermediate- and
high-mass stars.

The star-forming clumps lie outside of the average shockfront boundary of $251^{\prime \prime}$ ($4.6\,\rm pc$) from the centre but the nearest, faintest clump (Fig.~\ref{fig:multitycho}) is consistent with the distance reached by the outermost blast wave in this region at $260^{\prime \prime}$ (as traced by the hard X-rays).   Indeed, the travel-time for the shockfront to reach these star-forming cores
at radius $r \sim 5-6\,\rm pc$ from the centre is given
by $t = 2r/5v$ (assuming the SNR is in the Sedov phase where $v$ is
the expansion velocity - $v \sim 4700\,\rm km\,s^{-1}$ - Hayato et
al.\ 2010). The shockfront would therefore be expected to reach the observed
star-forming regions at $t= 425-510\,$yrs.   However, a simple timescale argument rules out any triggered
star formation by Tycho's SN since the gravitational collapse of a
cloud requires timescales of $10^5$\,yrs, with the proceeding
protostellar phase needing approximately $10^6$\,yrs. It is possible
that the stellar wind of the progenitor or the binary companion was
responsible for creating a low density bubble due to its stellar wind.
Such a scenario was suggested by Badenes et al.\ (2007) to explain
RCW86 (confirmed observationally by Williams et al.\ 2011) and the
star-forming loop around the SNR G54.1+0.3 (Koo et al.\ 2008).

\section{Discussion}
\label{sec:disc}

\subsection{A theoretical model of ejecta dust formation}
\label{sec:model}

Nozawa et al. (2011) have developed a theoretical model to follow the
dust formation in a carbon deflagration event (Nomoto, Thielmann \&
Yokoi et al. 1984) with ejecta mass of $1.38\rm \,M_{\odot}$. In this model, dust
grains form early-on, but are
almost completely destroyed over a timescale of $10^6$\,years due to
the passage through the reverse shock. The total dust masses formed at 100-300 days range from $3\times 10^{-4} - 0.2\,\rm M_{\odot}$, depending on the sticking probability of the grains and/or whether CO or SiO molecules are formed in the
ejecta (since this will affect the composition and mass of grains that condense out of the ejecta gas - Nozawa et al.\  2003).   Here, we use this model to compare the dust masses observed
with {\it Herschel} and the mass expected to have formed and survived
in the ejecta at approximately 400\,years after the explosion
event. (We refer the reader to the full description in Nozawa et al.\
(2011) for details of the simulation).  
    
\begin{figure}
\centering
\subfigure[]{\includegraphics[trim=1.5mm 0mm 5mm 0mm,clip=true,angle=0,width=9.0cm]{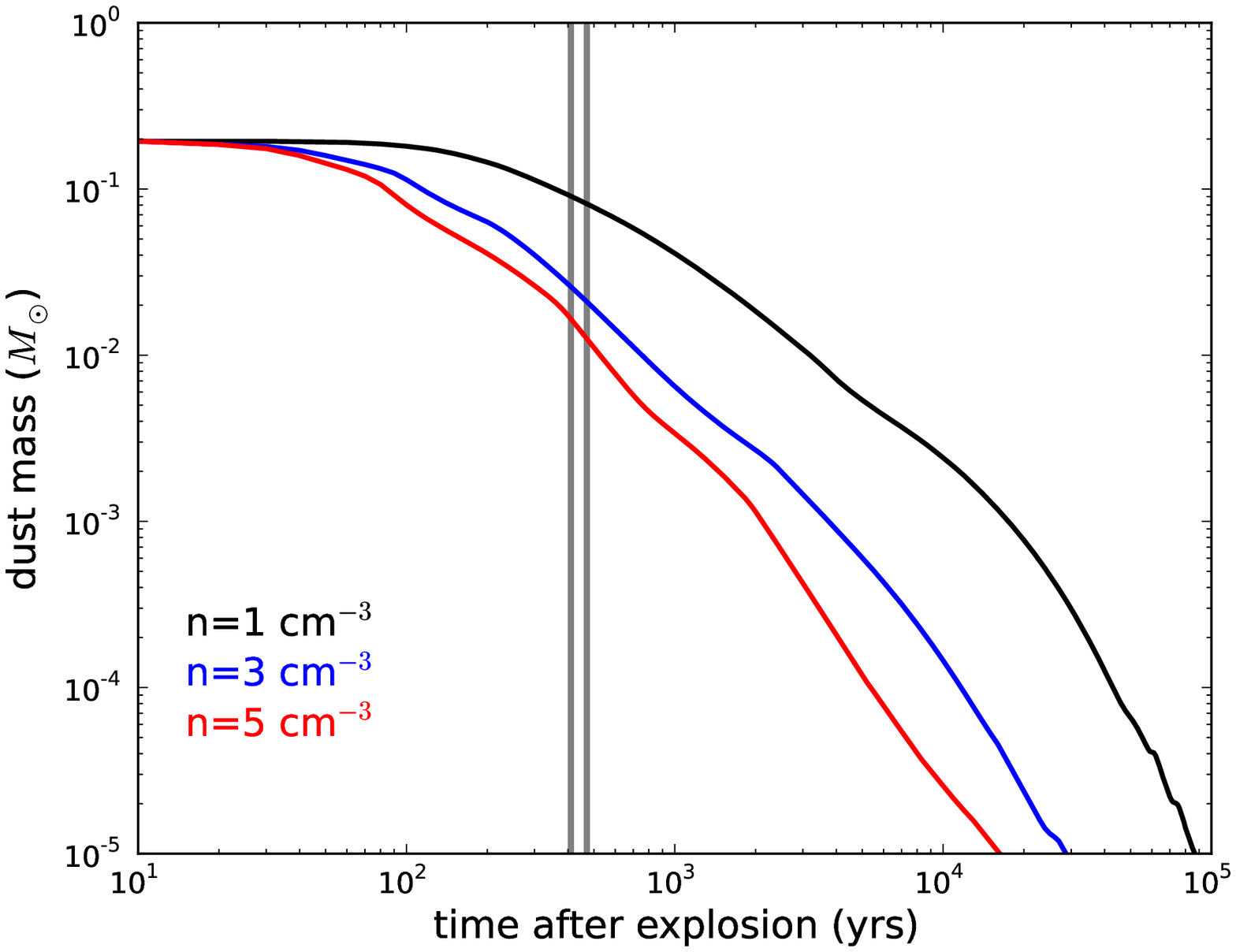}}
\subfigure[]{\includegraphics[trim=0mm 0mm 5mm 0mm,clip=true,angle=0,width=9.0cm]{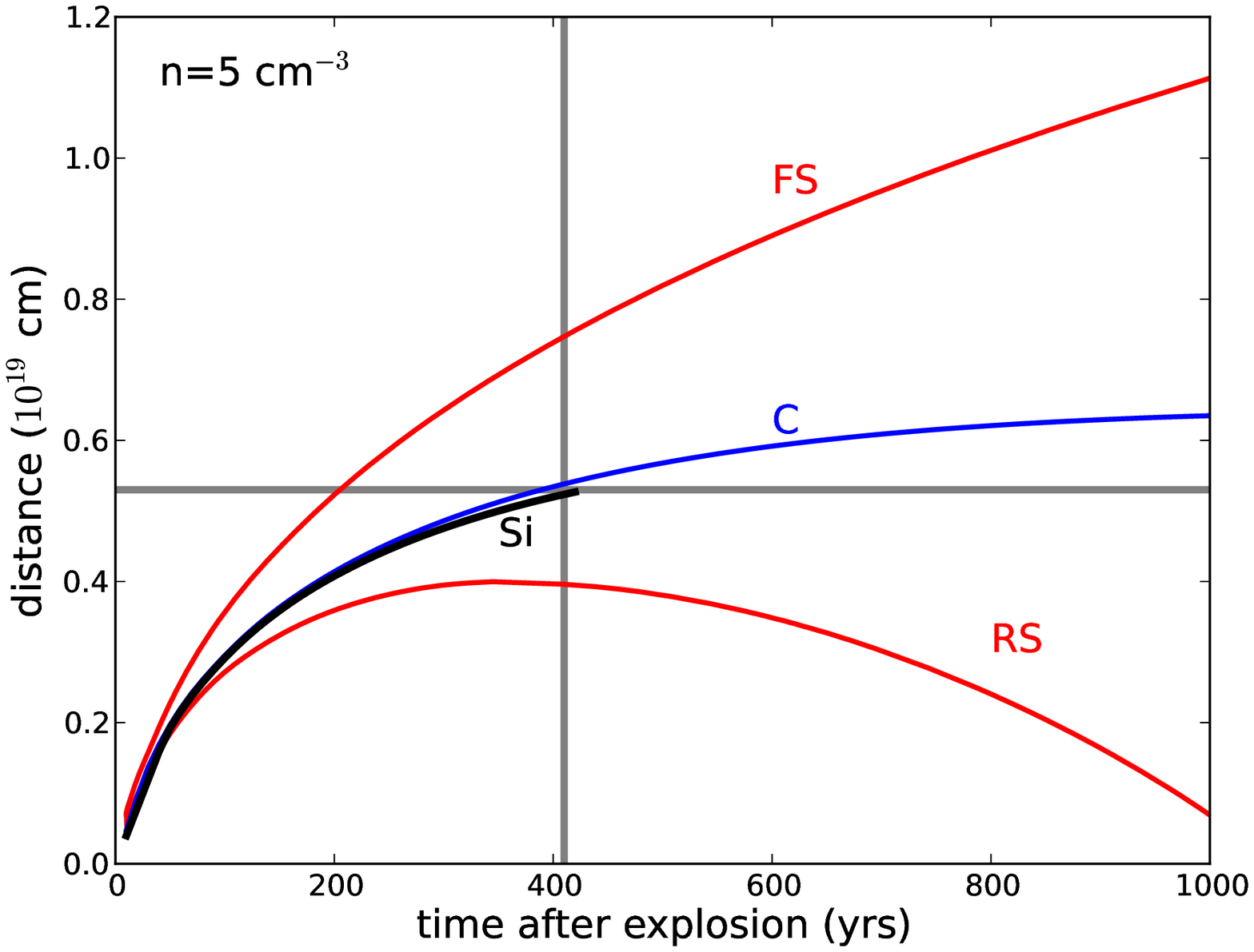}}
\caption{\small{{\bf (a)} The variation of dust mass with time since
    the supernova event in a Type Ia remnant based on the
    model of Nozawa et al.\ (2011). The dust mass with age are
    compared for different ambient densities of $n = 1\, \rm cm^{-3}$
    (black), $3\, \rm cm^{-3}$ (blue) and $5\, \rm cm^{-3}$ (red). The
    vertical lines indicate the time since explosion for Kepler and
    Tycho. {\bf (b)} The distance reached by ejecta dust with time
    since explosion for the Kepler SNR. The location of the forward
    (FS) and reverse shocks (RS) are indicated in red with the FS
    matched to the observed location at $110-120^{\prime \prime}$. The
    distance traveled by the dust grains are shown in blue (C) and black (Si). 
    The grey vertical line marks 410\,years and the horizontal line
    marks a distance of $5.3\times 10^{18}\,\rm cm$.  Note that the Si grains are destroyed just after 400\,years. }}
\label{fig:nozawa}
\end{figure}

Fig.~\ref{fig:nozawa} (a) shows the time evolution of the total dust mass of
supernova grains for a Type Ia using the dust formation calculation with a sticking probability equal to one.   We consider three ambient gas densities in which the ejecta is expanding into, where $n_{\rm H} = 1$,$3$ and $5\,\rm cm^{-3}$ since this will affect the amount of destruction of the dust grains we expect; the figure clearly shows that dust destruction is more efficient for higher ambient densities.
Using the relative ages of Kepler and Tycho at 410(440)\,years
and assuming that the ejecta is expanding into an ambient gas density of $1\,\rm cm^{-3}$, the model predicts a surviving
 dust mass of $\sim 88(84) \times 10^{-3}\, \rm M_{\odot}$:
an order of magnitude more than the warm dust mass of $3.1(8.6)\times
10^{-3}\,\rm M_{\odot}$ estimated from the {\it Herschel} observations (Table~\ref{tab:sed}).  If the
remnant is expanding into denser material, as certainly suggested for
both the Kepler and Tycho remnants, the surviving dust mass is
consequently reduced with $26(23)$ and $17(14) \times 10^{-3}\,\rm
M_{\odot}$ of dust surviving at 410(440) years for ambient densities
of $3$ and $5\,\rm cm^{-3}$ respectively; the surviving dust mass is still too high to be consistent with the observations.

The comparison between the dust mass estimated from simulations and that observed with {\it Herschel} implies that even if all the warm dust observed here was produced in the ejecta, the model still overestimates the amount of dust expected in the ejecta.  One possible explanation for this discrepancy is that the formation of dust grains in the ejecta
is less efficient than the model e.g. the sticking coefficient needed
to form the grains is lower, or that the radioactive heating from
nickel and cobalt is higher, therefore inhibiting the formation of dust (Nozawa et al.\ 2011). It is
also possible that dust destruction via sputtering may be more
efficient than the model, for example due to smaller grain sizes formed and/or higher ambient densities; the {\it Herschel} observations would then
suggest that dust destruction occurs efficiently on timescales of
$\sim$400\,years in Type Ia ejecta.  Finally, using a lower mass
absorption coefficient in our estimate of the observed dust masses would
also alleviate this discrepancy.  

Similarly,  we see no observational evidence for cool or cold dust associated with ejecta material as predicted using the dust formation model.  The lack of cold ejecta dust has important consequences on our understanding of the formation of iron-rich dust grains.  Type Ia ejecta are prime candidates for producing iron-rich dust as they synthesise an order of magnitude more $\rm ^{56}Ni$ (which eventually decays into Fe) than Type-II SNe.   Furthermore, there is a well known problem in reconciling the predicted ejecta abundances of iron using theoretical models (e.g. Thielemann, Nomoto \& Hashimoto 1996) with observed abundances which can differ by more than an order of magnitude.  It has long been suggested that one way to reconcile this issue is to `hide' the missing iron in dust grains.     The  Nozawa et al.\ (2011) model predicts that any newly formed iron-rich dust grains (which would form in the innermost regions of the ejecta) would not yet be swept by the reverse shock at 400\,days and would still be cold.  Since {\it Herschel} does not detect any significant ($\sim$subsolar) mass of cool dust associated with the ejecta, this strongly suggests that neither Kepler nor Tycho produced significant amounts of iron grains.    These observations suggest that the formation of dust  cannot explain the discrepancy between the observed and expected iron abundances in Type Ia supernova ejecta (see also Hamilton \& Fesen 1988).

\subsection{The origin of the warm dust component}
\label{sec:origins}

Given that we observe less dust than expected to be formed in the ejecta itself, could a non-supernova source of dust explain the IR emission?  One can estimate the mass of dust (Table~\ref{tab:sed}) we would
expect to be swept-up from circumstellar and interstellar material by
the SN blast wave using simple arguments. Kepler is not yet in the
Sedov phase (Decourchelle \& Ballet 1994; Kinugasa \& Tsunemi 1999;
Cassam-Chenai et al. 2004) which suggests that the remnant is still
dominated by the dynamics of the ejecta and has swept-up a mass
similar or less than that ejected in the explosion. The interstellar
density at the location of Kepler's remnant ($\sim 470\rm \,pc$ out of
the galactic plane) is expected to be low at $n_{\rm H} \sim
10^{-3}-0.5\,\rm cm^{-3}$ (Vink 2008). Estimating the volume encompassed by
the SN blast wave, the swept up interstellar gas mass would be $\sim
10^{-2}\,\rm M_{\odot}$ and hence the swept up dust mass is
$<10^{-4}\,\rm M_{\odot}$ (with gas-to-dust ratios of $100-150$ -
e.g. Devereux \& Young 1990). This is two orders of magnitude lower
than the dust mass observed in Kepler.

As suggested by Blair et al.\ (2007) and Douvion et al.\ (2001), the
origin for the IR emission in Kepler could be swept up circumstellar
dust. X-ray emission arises from this shocked dense circumstellar
component with mass $\sim 1\,\rm M_{\odot}$ (Kinugasa \& Tsunemi 1999;
Blair et al.\ 2007; Chiotelis, Schure \& Vink 2011). Given standard
gas-to-dust ratios, this mass could easily explain the warm dust
masses seen in the IR.  In this scenario, the explosion is
single-degenerate and the progenitor has a relatively massive evolved
companion, indeed Chiotellis, Schure \& Vink (2011) suggest the
companion was a $4-5\,\rm M_{\odot}$ AGB star.  Is the dust mass we
observe here consistent with the expected mass loss from an evolved
companion star?  Given the typical timescale for an AGB mass-loss
phase of $\sim 10^6\,\rm yr$ and typical dust mass-loss rates
of $\dot{M_{d}} \sim \rm 10^{-10}-10^{-8}\,M_{\odot}\,yr^{-1}$
(Meixner et al., 2004; Groenewegen et al. 2007; Lagadec et al. 2008), we
would expect $\rm \le 0.01\, M_{\odot}$ ejected during this
phase\footnote{Since some of the circumstellar/interstellar dust swept up by the supernova shock wave will also be destroyed via sputtering, the masses estimated from these simple swept-up volume arguments are therefore upper limits on the actual swept-up dust mass we'd expect.}. This is consistent with the
mass of warm IR-emitting dust in Kepler and supports either the
explosion of an intermediate mass WD-AGB binary system (e.g.
Chiotellis et al.\ 2011) or the single progenitor $8\,\rm M_{\odot}$
AGB-star thermonuclear explosion suggested by Reynolds et al. (2008).

In Tycho's remnant, it is unlikely that the swept-up dust originates
from the CSM as the companion star was inferred to be a low
mass sub-giant (Lu et al. 2011). Some of the dust could have been
created during the prior AGB phase of the WD star; using the argument
above, we could again expect to see $0.01\,\rm M_{\odot}$ of dust ejected in
this phase which is consistent with the measured dust mass for Tycho. However, there is no evidence for the presence of a dense CSM surrounding Tycho.
The ambient interstellar density as measured from X-ray and radio
evolution is often quoted as $n_{\rm H} <0.2\,\rm cm^{-3}$ (Dwarkadas
\& Chevalier 1998; Cassam-Chena\'{i} et al.\ 2004; Katsuda et al.\
2010) but recent simulations suggest $3\,\rm cm^{-3}$ (Badenes et al.\
2007).  We would therefore expect a swept-up dust mass$^6$ of up to
$0.013- 0.2\,\rm M_{\odot}$ for the range of ambient gas densities (with gas-to-dust ratio of 120). This is consistent with the warm dust component we
measure in Tycho's remnant.

\begin{table}
\centering
\begin{tabular}{llllll} \hline \hline
  & \multicolumn{3}{c}{Parameters} &\multicolumn{2}{c}{Dust mass}\\ 
  & \multicolumn{1}{c}{$\beta$}&\multicolumn{1}{c}{$T_{\rm w}$} & \multicolumn{1}{c}{$T_{\rm c}$} & \multicolumn{1}{c}{$M_{d,w}$} & \multicolumn{1}{c}{$M_{d,c}$}\\
  & \multicolumn{1}{c}{}&\multicolumn{1}{c}{  (K)
  } & \multicolumn{1}{c}{  (K)
   } & \multicolumn{1}{c}{$(\rm 10^{-3}\,M_{\odot})$} & \multicolumn{1}{c}{$(\rm M_{\odot})$}\\\hline 
Kepler & $1.5$ & $84$ & $19$ & $3.5$ & $2.2$\\ 
  &$1.6 \pm 0.2$ & $82^{+4}_{-6}$ & $18^{+2}_{-1}$ &
  $3.1^{+0.8}_{-0.6}$ & $2.1^{+0.8}_{-0.5}$\\   \hline
  Tycho &  $1.5\pm 0.2$ & $90$ & $21$ & $8.6$  & $3.7$ \\   
  &  $ 1.5 \pm  $ & $90^{+5}_{-7}$ &
  $21^{+2}_{-1}$ & $8.6^{+2.3}_{-1.8}$ & $5.1^{+0.9}_{-0.6}$ \\   \hline
\end{tabular}
\caption{\small{The best-fit parameters ({\it top row}) for dust
    towards and within Kepler and Tycho's SNRs and the
    median parameters ({\it bottom row})
    estimated from bootstrapping the fluxes within their
    errors.  We quote
    the median from the distribution of values obtained from the
    bootstrap analysis with the
    68\% per cent confidence interval quoted for the error.  The dust
    mass absorption coefficient, $\kappa$, used to estimate the warm dust
    mass in Kepler's and Tycho's SNR is $\kappa_{\rm 70\mu m}= 2\,\rm
    m^2\,kg^{-1}$.  For the cold dust component towards the SNRs, we
    use $\kappa_{\rm 500 \mu m} = 0.1 \,\rm m^2\,kg^{-1}$.  }}
\label{tab:sed}
\end{table}

Recently, Tian \& Leahy (2011), using H{\sc i} absorption lines, showed that a
number of the molecular clouds thought to be interacting with Tycho
are not at the location of the remnant. This would imply there is not enough interstellar medium surrounding Tycho for the remnant to sweep
up the dust we observe. However, they identify one H{\sc i} cloud which they rule out
as interacting with the remnant due to its density of $n_{\rm H} =
10\,\rm cm^{-3}$, yet we include this here since this is comparable to
the density of the surrounding material as estimated from
hydrodynamical modelling (Badenes et al.\ 2006). The mass of gas in
this cloud is $\sim \rm 10\,M_{\odot}$. Furthermore, we identify a
faint cloud overlapping with the warm dust in the NW of Tycho at
$-55\,\rm km\,s^{-1}$ (with gas mass $\sim 1\rm \,M_{\odot}$ using a
CO-$\rm H_2$ factor of $3\times 10^{-20}\rm
cm^{-2}\,(K\,km\,s^{-1})^{-1} $) which is massive enough to explain the origin of the
warm dust feature detected here$^6$ and in Ishihara et al.\ (2010).  

The warm dust we observe in these remnants is therefore coincident
with the outermost shockfronts, indicative of swept-up circumstellar
or interstellar material.   We can test this by investigating the
location of the dust grains expected to form from the ejecta gas using
the model of Nozawa et al.\ (\S\ref{sec:model}) w.r.t the forward and
reverse shocks. In Fig.~\ref{fig:nozawa} (b), we show the distance the
freshly-formed ejecta grains will have reached in 1000\,years with
ejecta ploughing into an ambient medium of density $n_{\rm H} = 5\,\rm
cm^{-3}$ (chosen so that the model forward shock location for Kepler
agrees with the observed shock radius at $\sim 110-120^{\prime
  \prime}$). This figure suggests that silicon and carbon dust grains
formed in the ejecta will encounter the reverse shock after 50\, years,
and at 400\,years (due to deceleration from the gas drag), will be
trapped at $5.3\times 10^{18}\,\rm cm$ from the centre; for $n_{\rm H}
= 10\,\rm cm^{-3}$, the dust would be trapped at $4.5\times
10^{18}\,\rm cm$. The model thus predicts that any supernova dust in
Kepler would lie at an angular distance of $78-91^{\prime
  \prime}$. The latter distance is consistent with the interior edge
of the dust emission seen with {\it Herschel}
(Fig.~\ref{fig:kepmulti}) but not consistent with the emission seen
out to $120^{\prime \prime}$. In Tycho, the observed warm dust is
confined between a thin region at $220-250^{\prime \prime}$ from the
centre (Figs.~\ref{fig:tychoco} and \ref{fig:tychoopt}). The model
suggests that for a forward shock of $\sim 250^{\prime \prime}$, the
supernova dust would be located at approx. $1\times 10^{19}\,\rm
cm$, i.e. at an angular distance of $200^{\prime \prime}$, again
consistent with the interior edge of the observed warm dust emission
in Tycho.

If the warm dust was formed in the supernova ejecta, we would expect
it to be distributed within the forward and reverse shocks across the
entire remnant. However, in Kepler's SNR, the warm dust exhibits a
north-south asymmetry, and in Tycho, there is a north-north-east
asymmetry and we therefore only observe warm dust emission where there
is also plenty of surrounding material. For Kepler, this is the
circumstellar material in the north which the remnant is ploughing
into as it moves at high velocities, and in Tycho, the dust originates
from regions where we see dense molecular clouds at the edge of the remnant.
This indicates that the majority of the warm dust is therefore swept-up. In
summary, we cannot absolutely rule out that a small fraction of the
warm dust in both remnants is formed in the supernova ejecta, but we
propose that, from the spatial distribution, the warm dust in Tycho
orginates from swept-up interstellar dust, and in Kepler, the warm
dust originates from swept-up circumstellar dust.

\subsection{The origin of the cold dust component}
\label{sec:kepcold}

Approximately $2-5\,\rm M_{\odot}$ of cold dust is found in the
apertures chosen to encompass the SNR, for Kepler and Tycho
respectively (Table~\ref{tab:sed}). The substantial amount rules out a swept-up CSM or ISM
origin of dust, but could this massive cold component be
freshly-formed dust in the supernova ejecta?

As well as using the temperature maps (Figs~\ref{fig:keptemp} \&
\ref{fig:tychotemp}), we also searched for an excess of submm emission
in the SNR compared to the unrelated molecular gas. The CO and submm
fluxes (using the 250\,$\mu$m) in each pixel were compared and a
best-fit relation determined, producing, for each pixel, an estimate
of the 250\,$\mu$m flux that corresponds to a given CO flux. The
scatter in the CO and 250-$\mu$m fluxes for Kepler's remnant is
considerable, and when the submm flux associated with the CO is
subtracted from the 250\,$\mu$m map, residual submm emission remains
in the northern shell (in the same location as the structure seen in
the cold temperature map, Fig~\ref{fig:keptemp} (b)); the dust mass in
this feature is $\sim 0.6\rm \,M_{\odot}$ for $T_d\sim \rm 20\,K$.  The
cold dust does not spatially coincide with any of the ejecta tracers,
and therefore we conclude that the submm emission is not dust formed in
the supernova ejecta.

There is no obvious CO structure at the location of the cold dust
shell seen in Kepler, though the difference between the flux at
250\,$\mu$m in this region and the interstellar cloud to the east of
the remnant is a factor of 5; the expected CO emission associated with
this faint cold dust feature is therefore at the noise level of the CO
map. There is submm `excess' over the molecular gas both inside and
outside the remnant, suggesting that these dust clouds in general are
not always well traced by the $^{12}$CO(J=2-1) emission; this method
is therefore not able to distinguish between dust associated with the
foreground dust component and the remnant.

Towards Tycho, a large mass of cold dust exists within the aperture
encompassing the forward blast wave. Assuming that this dust component
originates from the swept-up interstellar material would imply more
than $600\,\rm M_{\odot}$ of gas has been swept up in the last
440\,years. This is unphysical since Tycho is still in the Sedov phase
where the swept up mass $\le 10\times M_{\rm ejecta}$. The cold dust
on the eastern edge of the remnant as seen clearly in the temperature
map (Fig.~\ref{fig:tychotemp}) appears (at first glance) to be
swept-up cold dust, but the large mass of dust in this feature ($\sim
2\,\rm M_{\odot}$) again rules this scenario out. There are a
significant number of atomic and molecular clouds overlapping on the
line of sight, the average $\rm H_2$ column density towards the
remnant (as estimated from the CO map) is $\sim 10^{21}\,\rm cm^{-2}$,
with clouds of sizes $3-6\rm \,pc$ at distances of $3.4-5\,\rm kpc$,
suggesting gas masses of $\sim 500-2000\,\rm M_{\odot}$. Therefore,
given the abundance of gas towards and behind Tycho (and spatially
bounded between the SN shock radius), it is likely that we are
observing a small amount of cold dust swept up in the ISM along with a
significant amount of dust emission from interstellar clouds along the
line of sight to the remnant.

\section{Conclusions}
\label{sec:conc}

{\it Herschel} observations of the remnants of the core-collapse SNe Cas A and SN1987A show evidence for a significant mass of dust
spatially coincident with SNe ejecta. Using {\it Herschel}
observations of Kepler and the Type Ia SNR, Tycho, we do not find
evidence for freshly-formed supernova dust in these historical
remnants.

We confirm a previously known warm dust component in Kepler's remnant
with $T_d =82^{+4}_{-6} \, \rm K$, somewhat cooler than the dust
previously detected by {\it Spitzer} and {\it ISO}, and revise the
dust mass to $\rm \sim (3.1^{+0.8}_{-0.6})\times 10^{-3}
\,M_\odot$. The warm dust is spatially coincident with thermal X-ray
emission and optical knots and filaments on the outer edges of the
shockfront, confirming that this component originates in the material
swept up by the primary blast-wave. Given the mass and the density of
the surrounding ISM, the warm dust could not have originated from
swept up interstellar material, but is consistent with swept-up {\it
  circumstellar} dust, confirming the presence of CSM surrounding
Kepler. The {\it Herschel} SPIRE images are dominated by emission from
cold foreground insterstellar clouds ($\sim 20\,\rm K$) which overlap
with the small dust clumps detected by previous ground-based SCUBA
measurements. These were originally attributed to a large mass of dust
either formed in the ejecta and/or swept up by the circumstellar
medium. The resolution of {\it Herschel} does not allow us to
distinguish between cold supernova dust or intervening interstellar
material, however using dust temperature maps, and a comparison with
molecular gas (traced by CO), we find that most of the cold dust
originates from foreground, unrelated ISM. There is no evidence that
any of the cold dust within the aperture is spatially coincident with
ejecta material.

Similarly for Tycho's remnant, we detect warm dust at
$90^{+5}_{-7}\rm \,K$ with mass $(8.6^{+2.3}_{-1.8})\times
10^{-3}\,\rm M_{\odot}$.  Comparing the spatial distribution of the
warm dust with thermal and non-thermal X-rays from the ejecta and swept
up medium, and with collisionally heated H$\alpha$ emission arising
from the thin post-shock edge, we suggest the warm dust is swept up
{\it interstellar} material. This is further supported by the
significant amount of molecular gas surrounding the regions where warm
dust is observed and the large mass of interstellar gas swept up by the
remnant.  We find no evidence for cool supernova dust at temperatures
of $20-30\,\rm K$ as detected in the core collapse remnants Cas A and
SN1987A; the lack of freshly formed dust in both
remnants supports the view that Kepler was the result of a Type Ia explosion.

We compare the dust masses with the recent theoretical dust formation
model of Nozawa et al.\ (2011) where approximately $\rm
10^{-2}\,M_{\odot}$ of dust is expected to form in Kepler and Tycho.
The Si and C grains formed are expected to have traversed the reverse
shock approximately 50\,years after the explosion and are then trapped
at 400\,years at a distance consistent with the interior edge of the
warm dust emission detected by {\it Herschel}.   The observed dust mass of the
warm component is lower than the model predicts, suggesting that dust
formation in the ejecta is inhibited or destroyed more efficiently.    We do not detect a cool dust component in the innermost regions of un-shocked ejecta, suggesting that Type Ia ejecta are not producing substantial masses of iron and are not responsible for the `missing' iron mass problem in the ejecta.  The mass of dust in these remnants is lower than that measured in
core-collapse SNRs, though there is a possibility that a fraction of
the cold dust observed towards both Tycho and Kepler could be hidden
amongst the dust components along the line of sight. Resolved
observations of younger Type Ia remnants with low
ambient densities (and little foreground contamination) are needed to
investigate the evolution of
  supernova dust mass further.

\section*{Acknowledgements} We thank the referee for their helpful comments on the manuscript.  The images were produced with the software package APLpy {\em http://aplpy.github.com/}: thanks to Eli Bressart
\& Thomas Robotaille for this. We thank Jason Kirk, Tracey DeLaney and Chris North for useful
discussions. H.L.G.\ would like to acknowledge the support of Las
Cumbres Observatory Global Telescope Network. 

PACS has been developed by a consortium of institutes led by MPE
(Germany) and including UVIE (Austria); KU Leuven, CSL, IMEC
(Belgium); CEA, LAM (France); MPIA (Germany); INAF-IFSI/OAA/OAP/OAT,
LENS, SISSA (Italy); IAC (Spain). This development has been supported
by the funding agencies BMVIT (Austria), ESA-PRODEX (Belgium),
CEA/CNES (France), DLR (Germany), ASI/INAF (Italy), and CICYT/MCYT
(Spain). SPIRE has been developed by a consortium of institutes led by
Cardiff University (UK) and including Univ. Lethbridge (Canada); NAOC
(China); CEA, OAMP (France); IFSI, Univ. Padua (Italy); IAC (Spain);
Stockholm Observatory (Sweden); Imperial College London, RAL,
UCL-MSSL, UKATC, Univ. Sussex (UK); and Caltech/JPL, IPAC, Univ.
Colorado (USA). This development has been supported by national
funding agencies: CSA (Canada); NAOC (China); CEA, CNES, CNRS
(France); ASI (Italy); MCINN (Spain); Stockholm Observatory (Sweden);
STFC, UKSA (UK) and NASA (USA). This work is based [in part] on
observations made with the Spitzer Space Telescope, which is operated
by the Jet Propulsion Laboratory, California Institute of Technology
under a contract with NASA.

\end{document}